\documentclass[%
superscriptaddress,
preprint,
 amsmath,amssymb,
 aps,
]{revtex4-2}

\usepackage{graphicx}
\usepackage{dcolumn}
\usepackage{bm}

\usepackage{xcolor}
\usepackage{newtxtext,newtxmath}

\newcommand{\edit}[1]{{\color{black} #1}}

\begin{document}

\preprint{APS/Physical Review Fluids}

\title{Deep learning for surrogate modeling of two-dimensional mantle convection}

\author{Siddhant Agarwal}\thanks{Funded by HEIBRiDS Graduate School for Data Science}
 \email{agsiddhant@gmail.com}
 \affiliation{Planetary Physics, Institute of Planetary Research, German Aerospace Center (DLR), Berlin, Germany}
 \affiliation{Machine Learning Group, Berlin Institute of Technology, Berlin, Germany}
 
\author{Nicola Tosi}
\affiliation{Planetary Physics, Institute of Planetary Research, German Aerospace Center (DLR), Berlin, Germany}

\author{Pan Kessel}\thanks{Funded by BIFOLD}
\affiliation{Machine Learning Group, Berlin Institute of Technology, Berlin, Germany} 

\author{Doris Breuer}
\affiliation{Planetary Physics, Institute of Planetary Research, German Aerospace Center (DLR), Berlin, Germany}

\author{Gr\'egoire Montavon}\thanks{Funded by BIFOLD}%
\affiliation{Machine Learning Group, Berlin Institute of Technology, Berlin, Germany}

\date{\today}

\begin{abstract}
Mantle convection, the buoyancy-driven creeping flow of silicate rocks in the interior of terrestrial planets like Earth, Mars, Mercury and Venus, plays a fundamental role in the long-term thermal evolution of these bodies. Yet, key parameters and initial conditions of the partial differential equations governing mantle convection are poorly constrained. This often requires a large sampling of the parameter space to determine which combinations can satisfy certain observational constraints. Traditionally, 1D models based on scaling laws used to parameterized convective heat transfer, have been used to tackle the computational bottleneck of high-fidelity forward runs in 2D or 3D. However, these are limited in the amount of physics they can model (e.g. depth dependent material properties) and predict only mean quantities such as the mean mantle temperature. A recent machine learning study has shown that feedforward neural networks (FNN) trained using a large number of 2D simulations can overcome this limitation and reliably predict the evolution of entire 1D laterally-averaged temperature profile in time for complex models. We now extend that approach to predict the full 2D temperature field, which contains more information in the form of convection structures such as hot plumes and cold downwellings. Using a dataset of $10,525$ two-dimensional simulations of the thermal evolution of the mantle of a Mars-like planet, we show that deep learning techniques can produce reliable parameterized  surrogates (i.e. surrogates that predict state variables such as temperature based only on parameters) of the underlying partial differential equations. We first use convolutional autoencoders to compress the \edit{size of each} temperature field by a factor of $142$ and then use FNN and long-short term memory networks (LSTM) to predict the compressed fields. On average, the FNN predictions are $99.30\%$ and the LSTM predictions are $99.22\%$ accurate with respect to unseen simulations. Proper orthogonal decomposition (POD) of the LSTM and FNN predictions shows that despite a lower \edit{mean relative accuracy}, LSTMs capture the flow dynamics better than FNNs. When summed, the POD coefficients from FNN predictions and from LSTM predictions amount to $96.51\%$ and $97.66\%$ relative to the coefficients of the original simulations, respectively. 

\end{abstract}

\maketitle


\section{Introduction}

Studying the long-term thermal evolution of terrestrial planets like Earth, Venus, Mercury and Mars requires detailed modelling of thermal convection in their rocky mantles (e.g., \cite{breuer2015}). Similar to the flow of crystalline ice in glaciers, mantle convection is a form of sub-solidus convection. Mantle rocks at high temperature and pressure, but still well below their melting point, behave like a highly viscous fluid over geological time scales (millions to billions of years), largely in response to thermal and compositional buoyancy forces. Mantle convection is typically modelled using fluid dynamics codes (e.g., \cite{Tackley2008,Zhong2008,Kronbichler2012,huettig2013}) that numerically solve the non-linear partial differential equations (PDEs) of mass, momentum and energy conservation governing the creeping flow (i.e. with negligible inertia) of silicate rocks subject to basal heating from the metallic core and internal heating due to the decay of radioactive elements. Unfortunately, key model parameters (such as the rock viscosity, or the amount of internal heating) and initial conditions (such as the initial mantle and core temperatures), that are inputs to these forward numerical models, are poorly constrained. Thus, one typically chooses and tests a large number of different parameter values (within a reasonable range) to observe how these affect the outputs of the simulations. The outputs can be processed to arrive at various quantities of interest such as surface heat flux, amount of thermal contraction or expansion, duration and timing of volcanism, or volume of produced crustal material. To some extent, these quantities are ``observables'' that can be inferred from geophysical and geochemical data delivered by planetary space missions. In turn, they provide fundamental constraints on the convective evolution of terrestrial bodies (see \cite{Tosi2021} for a recent review about this topic).   

This approach of testing several different values of parameters, however, suffers from a bottleneck imposed by the computational cost of the 2D and 3D forward models. No matter whether one approaches this issue from the perspective of an inverse problem (inferring the model parameters given the observables) or of a forward problem (calculating the observables given the model parameters), it is often impractical to run several thousands of simulations to determine which parameters and combinations thereof can satisfy a set of given observational constraints. 

A number of inverse-problem studies have attempted to overcome this computational bottleneck of expensive simulations, ranging from using modified Markov Chain Monte Carlo (MCMC) methods \edit{(e.g. \cite{Sambridge_a, Sambridge_b})}, all the way to completely bypassing MCMC methods and directly learning the mapping between parameters and observables from simulations run prior to the inversion using Mixture Density Networks (MDN) (e.g., \cite{Meier2007,Kauefl2016,deWit,atkins,agarwal2021}).

Machine learning (ML) methods such as MDN can also be used to learn highly non-linear forward mappings from parameters to observables. These can preserve some physical insights into the flow being modelled (such as convection patterns) in contrast to purely statistical inferences made under an inverse formulation. Traditionally, several mantle convection studies have employed ``scaling laws'', which parameterize the heat flow out of the mantle (quantified by the Nusselt number ($Nu$), or the non-dimensional heat flux), in terms of the vigor of convection (quantified by the Rayleigh number ($Ra$)) (e.g., \cite{reese1998,dumoulin1999,solomatov2000,deschamps2001}). These scaling laws are then used in the frame of one-dimensional, spherically-symmetric models to advance the mean mantle temperature (and a number of associated quantities) in time by solving two ordinary differential equations governing the global energy balance of the mantle and core (e.g., \cite{stevenson1983,gurnis1989,schubert1990,hauck2004,korenaga2011,morschhauser2011,tosi2013b,orourke2015}). 

On the one hand, the use of scaling laws makes models of planetary evolution computationally very efficient as these only require the solution of ordinary differential equations. On the other hand, scaling laws are limited in the amount of physics they can capture. For example, solid-solid phase-transitions or the pressure-dependence of the viscosity and other thermal and transport properties such as thermal expansivity and conductivity can hardly be taken into account in scaling laws for heat transfer. Additionally, they allow one only to predict the evolution of global quantities such as the surface heat flux or the mean mantle temperature, but do not provide any insight into the spatial and temporal variability of mantle flow. Recently, \cite{Shahnas2020} showed that some of these limitations can be overcome by using machine learning. They showed that feedforward neural networks (FNN) can be used for predicting the surface heat flux and mean temperature of steady-state simulations, given parameters such as $Ra$, the core-to-planet radius ratio, and mode of convection, i.e. mobile lid, characterizing  Earth's plate tectonics, or stagnant lid, characterizing the other terrestrial bodies of the solar system (see also Sec. \ref{sec-data}). \cite{agarwal2020} demonstrated that limitations related to taking into account complex physics and the time-variability of the heat transfer can be overcome through FNNs properly trained with a large set of 2D dynamic simulations. By using multiple parameters such as e.g. mantle reference viscosity (related to $Ra$) and activation volume and activation energy of diffusion creep rheology (controlling the temperature and pressure dependence of the viscosity) as inputs to the FNN, it is possible to directly predict the thermal evolution of the entire horizontally-averaged 1D temperature profile of the mantle in time with a mean accuracy of $99.7\%$. \edit{Another interesting study uses Generative Adversarial Networks to reconstruct missing plate boundaries derived from horizontal divergence maps of a steady-state 3D convection model \cite{Gillooly2019}.}

\begin{figure*}
\centering
\includegraphics[width=\textwidth]{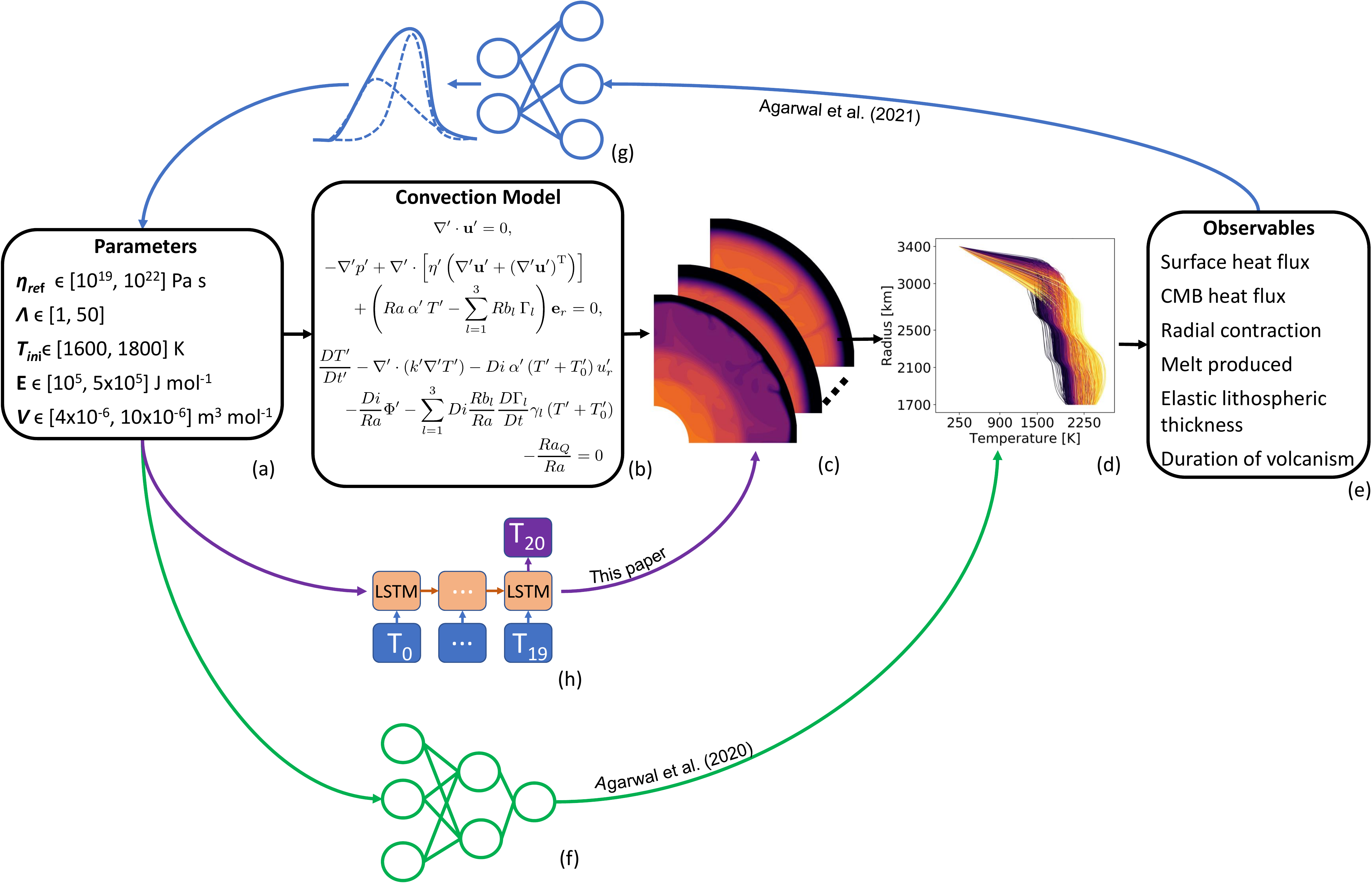}
\caption{The context for this study. (a) Typically, a mantle convection study starts by randomly drawing input parameters from a flat distribution and then feeding them to the forward models. (b) The PDEs are solved using dedicated mantle convection codes. (c) The outputs of the simulations can be processed to arrive at certain lower-dimensional observables such as (d) the horizontally-averaged 1D temperature profiles or (e) more global quantities such as the surface heat flux, radial contraction, duration of volcanism, etc. ML methods have been shown to work well for these low-dimensional observables, both - (f) in a forward study (\cite{agarwal2020}) and (g) an inverse study (\cite{agarwal2021}). In this work, we demonstrate that (h) a surrogate can model 2D mantle convection using deep learning. }
\label{fig-concept}
\end{figure*}

In this paper, we build upon \cite{agarwal2020}. While the full 1D temperature profile already provides a lot more information than simply the surface heat flux and mean temperature and is also beyond what can be expected to be retrieved for planets like Mars through future planetary missions, it still lacks the rich convection structures such as plumes and downwellings that are delivered by 2D or 3D simulations. In fact, \cite{agarwal2021} demonstrated that using the same setup for a Mars-like planet as in \cite{agarwal2020}, the surface heat flux provides only a loose constraint for parameters governing mantle convection and even knowledge of the entire 1D temperature profile is insufficient for constraining certain parameters (e.g. the activation volume of diffusion creep rheology, which controls the pressure dependence of the viscosity). In other words, more information (such as horizontal variations of the temperature field) might hold the key to placing tighter constraints on certain thermal-evolution parameters. Hence, it is desirable to construct surrogate models capable of predicting more information about a convecting mantle. 

Here, we show how deep learning (DL) can be leveraged to directly predict surrogates of the 2D temperature fields from five key-parameters. Figure \ref{fig-concept} shows the concept and general context within which this study fits. We sample the model parameters from a flat distribution with a broad, yet reasonable range and input these to the convection simulations. The five key parameters are: (1) reference mantle viscosity ($\eta_{\rm ref}$), which is the viscosity attained at a given reference depth and reference temperature; (2) activation energy of the diffusion creep ($E$), which controls the temperature dependence of the viscosity, and (3) activation volume of diffusion creep ($V$) controlling the degree to which viscosity depends on pressure; (4) enrichment factor ($\Lambda$), which determines the proportion of the heat-producing radiogenic elements extracted from the convecting mantle upon melting and enriched in the crust; (5) the initial temperature of the mantle ($T_{\rm ini}$). \edit{
These five parameters typically have the largest influence on the forward model for the thermal evolution of terrestrial planets that we use (e.g. \cite{plesa}, \cite{plesa2018}).}

To our knowledge, this is the first time a parameterized 2D surrogate is proposed in the context of mantle convection (see \cite{Morra2020} for a recent review of applications of data science methods in geodynamics). We use the term parameterized to stress that we are interested in predicting a variety of mantle flows based on different combinations of input parameters and not, for example, given a certain amount of time-steps of a single simulation with fixed parameters, predicting the subsequent time-steps. \cite{DNSConvect}  demonstrated that direct numerical simulation (DNS) of two-dimensional turbulent Rayleigh-B\'enard convection can be modelled using reservoir computing. In the above study, the time-steps of the same, single simulation are split into training and test sets. In contrast, we are interested in modelling all the time-steps of different simulations, which is made possible by the relatively low computational cost of each simulation ($20$ to $500$ CPU hours depending on the combination of parameters).

In general, machine learning for predicting flows is an active research area. Particularly worth highlighting is the seminal paper by \cite{pinn}, which showed that the PDEs can be embedded in the loss function through automatic differentiation of state variables with respect to spatial and temporal coordinates. However, here we stick to a purely data-driven approach, reserving the application of methods such as those in \cite{pinn} for future research. A notable example of such a purely data-driven approach was shown to be effective in capturing the dynamics of 3D turbulence by \cite{Mohan2020}. They showed that the velocity fields can be compressed using convolutional autoencoders \cite{convae}. \edit{Convolutional autoencoders successively down-size the original field (or image) into a bottleneck, from where they are reconstructed back to the original size. In this way, the dimensionality of the original high-resolution fields can be decreased and made more computationally efficient to work with. \cite{Mohan2020} then predicted these compressed time-steps using a convolutional long short-term memory (LSTM) network \cite{convlstm}, which by providing a mechanism to relate the time-steps of a simulation, allows one to learn the attractor for the underlying dynamics.} Just as \cite{DNSConvect}, \cite{Mohan2020} split the time-steps of the same simulation into training and test sets. We show that this approach can be adapted to our purposes, so that we can predict all the time-steps for any simulation in our data manifold, given just a set of five parameters. Somewhat closer to our purposes, \cite{duraisamy_2019} used a convolutional encoder-decoder architecture for predicting pressure and temperature fields around airfoils, given the spatial grid as well as two additional parameters (angle of attack and Reynolds number).  We refer to \cite{BruntonMLReview} for an overview of several ML techniques that have been used for prediction, dimensionality reduction and optimization and control in fluid dynamics.

The outline of the paper is as follows. In the Sec. \ref{sec-data}, we briefly outline the setup of the simulations of a Mars-like planet. Then, in Sec. \ref{sec-compress}, we present how we compress the temperature fields obtained from the mantle convection simulations. We follow up with Sec. \ref{sec-predict}, where we predict the compressed temperature fields using two different ML algorithms - FNN and LSTM. In the same section, we further delve into the differences in the predictions of the two algorithms by analyzing them from the lens of Proper Orthogonal Decomposition (POD) or its ML equivalent, Principal Component Analysis. We then conclude the paper by offering some potentially interesting follow-ups to this work.

\section{Dataset of mantle convection simulations}
\label{sec-data}

We employ a dataset consisting of $10,525$ simulations of the thermal evolution of a Mars-like planet run on a 2D quarter-cylindrical grid (Fig. \ref{fig-sketch}). A detailed description of the setup is provided in Appendix \ref{Section_Appendix_MC}. Here, we summarize the main features of the model. 

We model the mantle as a viscous fluid with negligible inertia (i.e. with infinite Prandtl number or, equivalently, undergoing Stokes flow). We consider a pressure- and temperature-dependent Newtonian rheology \cite{arrhenius}, which is calculated using the Arrhenius law for diffusion creep\edit{, whose dimensional form reads
\begin{equation}
    \eta(T,P) = \eta_{\rm ref} \exp \left(\frac{E+PV}{T} - \frac{E+P_{\rm ref}V}{T_{\rm ref}} \right). \label{eq:arrhenius}
\end{equation}
The reference viscosity $\eta_{\rm ref}$ is attained at reference temperature $T_{\rm ref}=1600$ K and reference pressure  $P_{\rm ref}=3$ GPa, respectively. $P$ is the hydrostatic pressure, $E$ is the activation energy, and $V$ is the activation volume.}

Mars, in contrast to the Earth, but like Mercury, the Moon and, at least at present-day, Venus, operates in a so-called stagnant-lid convection mode (see e.g. \cite{Tosi2021}). The strong temperature dependence of the viscosity causes the relatively cold upper part of the mantle to develop a high-viscosity layer -- the stagnant lid. Such a stiff layer remains immobile during the entire evolution of the planet (although its thickness can decrease or increase in response to heating or cooling of the mantle). The stagnant lid insulates the interior causing thermal convection to take place only in the mantle beneath it and to be largely driven by cold downwellings developing at its base (see Fig. \ref{fig-sketch}). This mode of convection is remarkably different from the plate tectonic (or mobile-lid) mode of convection that characterizes the Earth. Cold tectonic plates are in fact an active part of the Earth's convecting engine: by sinking into the mantle at subduction zones, they provide a strong cooling for the mantle and core with fundamental consequences for large-scale transport of materials in the deep interior and for the generation of the Earth's magnetic field. 

\begin{figure*}[ht!]
\centering
\includegraphics[width=0.8\textwidth]{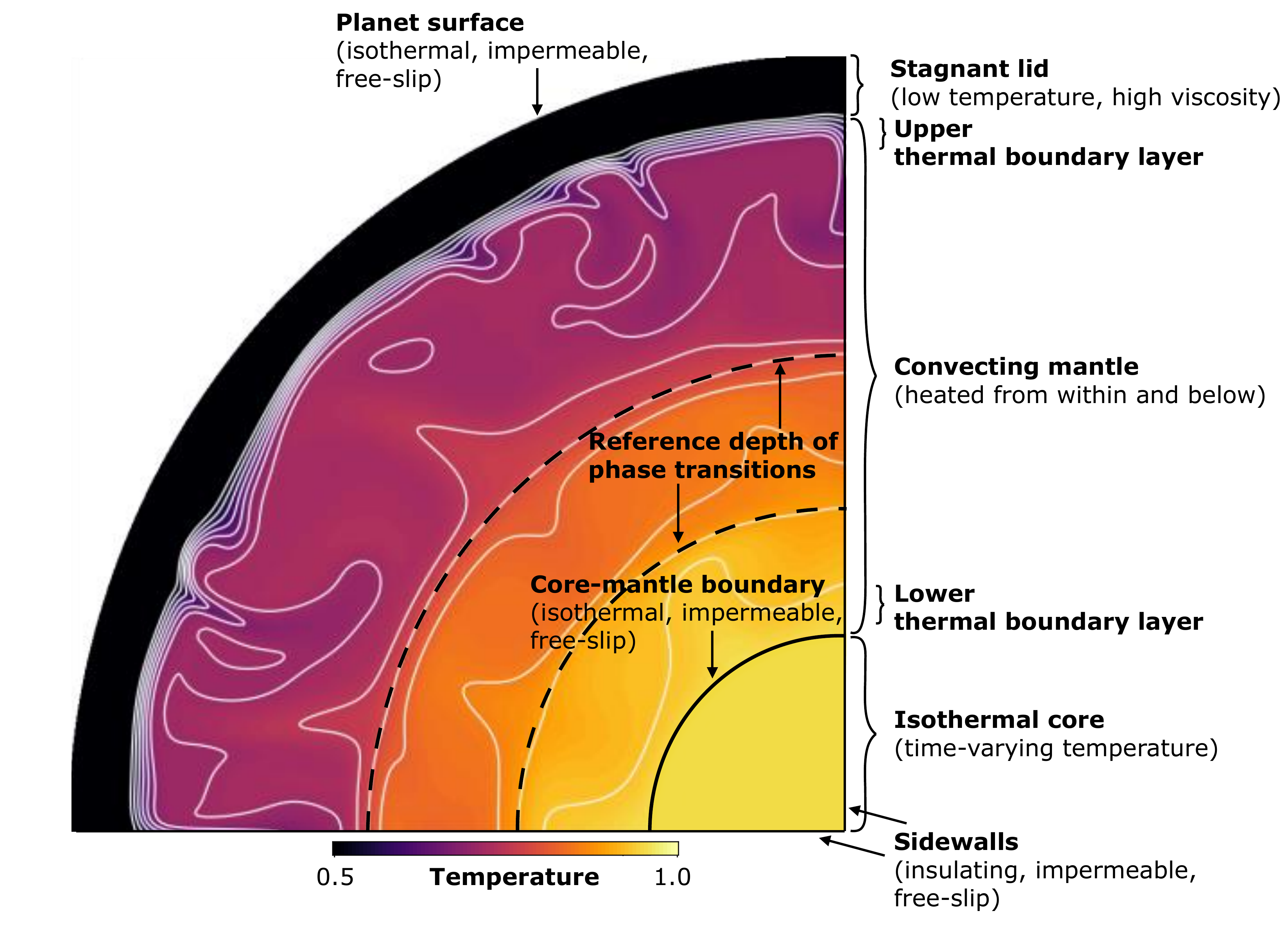}
\caption{Two-dimensional, quarter-cylindrical domain illustrating the main features of the employed mantle convection model. The mantle is colored according to the non-dimensional temperature field of one specific simulation of the dataset. The diagram depicts the smaller rescaled core (see Sec. \ref{ref_subsection_coreRescale}). Since a large part of the temperature variations occur across the stagnant-lid, the colorscale is truncated at 0.5 for ease of visualization. See text for more details and Appendix \ref{Section_Appendix_MC} for a complete model description.}
\label{fig-sketch}
\end{figure*}

Within the relatively low pressure range of Mars mantle ($\sim 20$ GPa), the degree of compressibility of mantle rocks is limited; the dissipation number (eq. \ref{eq-Di}) is quite small for Mars ($\sim 0.13$, significantly smaller than the Earth's, which amounts to $\sim 0.5$). In this situation, it is appropriate to employ the so-called extended Boussinesq approximation (e.g., \cite {king2010}). Like the standard Boussinesq approximation, the extended one assumes constant density everywhere except in the buoyancy term of the momentum equation (eq. \ref{eq-EBA2}). However, it further accounts for the effects of adiabatic compression/decompression and viscous dissipation in the conservation equation for the thermal energy (eq. \ref{eq-EBA3}).

The mantle temperature evolves in time due to the decay of radiogenic elements present in the mantle and due to the cooling of the core, which also provides a source of basal heat. Following a standard approach adopted in the mantle convection community, the core is simply considered as a homogeneous sphere of given density and heat capacity, whose mean temperature evolves at a rate imposed by the cooling rate of the mantle (see Fig. \ref{fig-sketch}, eq. \ref{eq-CoreCool} and e.g., \cite{stevenson1983}).

Geological evidence suggests that the bulk of the crust of Mars formed very early in the evolution of the planet \cite{Nimmo}. Similar to other studies (e.g., \cite{plesa2018}), we thus assume that a crust of a fixed thickness has been present since the beginning of the evolution. This assumption implies that, from the beginning of each simulations, the mantle is partly depleted of radiogenic elements (resulting in new new depleted mantle composition $C_{\rm depleted}$) with respect to their primordial \edit{concentration ($C_0$), which we set according to the model of \cite{waenke1994} as follows:
\begin{equation}
\label{eq-HP1-main}
C_{\rm depleted} = \frac{M_m C_0}{M_{\rm cr} \left (\Lambda - 1 \right )  + M_{\rm m}},
\end{equation}
where $M_{\rm m}$ and $M_{\rm cr}$ are the mass of the mantle and crust, respectively and $\Lambda$ is the crustal enrichment factor.} The rationale of this assumption is that upon partial melting of the mantle -- a common event during the evolution of terrestrial bodies -- radiogenic elements behave as incompatible elements, i.e. they tend to be enriched in the melt phase with respect to the solid phase. As a consequence, the production of crust, which results from mantle melting, melt migration toward the surface and solidification, causes a net depletion of  radiogenic elements in the mantle. In practice, we reduce the primordial bulk abundances of radiogenic elements according to a crustal enrichment factor $\Lambda$. The mantle is further depleted in radiogenic heat-producing elements during the evolution whenever local partial melting takes place according to the model described by \cite{padovan} (see Sec. \ref{sec_melting}). 

We also consider the influence on the mantle flow of two major solid-solid phase-transitions using the standard phase-function approach of \cite{phase} adopted in mantle convection (Sec. \ref{sec_phasetrans}). We further assume that the coefficient of thermal expansion and the thermal conductivity depend on pressure and temperature as appropriate for silicate materials (see Sec. \ref{sec_alphak} and \cite{tosi2013a}). 

The simulations are initialized with a constant mantle temperature $T_{\rm ini}$ combined with upper and lower $300$-km-thick thermal boundary layers. For all simulations, a small random perturbation is added to the temperature field to initiate convection. \edit{The initial mantle temperature has a strong effect on the early evolution of the planet. However, this initial condition becomes less important with time (after $\sim 2$ billion years) due to the ``thermostat effect'' \cite{tozer1967} - the temperature dependence of the viscosity regulates the temperature of the mantle. On the one hand, when the mantle is hot, the viscosity decreases, leading to more vigorous convection and thereby, efficient cooling. On the other hand, when the mantle is cooler at latter stages in the evolution, it is more viscous and cools less efficiently. While in this dataset of forward models we only vary the initial mantle temperature, other parameters related to the initial conditions that can have a potential impact on the thermal evolution of the interior could also be considered. For example, the starting temperature at the core-mantle boundary has implications for the dynamics of the lower mantle and affects melt and magnetic field generation (e.g. \cite{breuer2003}). The initial thickness of the thermal boundary layers, although irrelevant for the long-term evolution, influences the heat fluxes across the surface and CMB during the first few hundred million years. Additionally, an initial compositional stratification resulting from the crystallization of a primordial magma ocean, neglected in our isochemical models, could also have a significant influence on the evolution of the mantle and core (\cite {tosi2013c,plesa2014, scheinberg2014}. }

As for the boundary conditions, all domain boundaries are impermeable and free-slip. The surface temperature is kept fixed at $250$ K throughout the evolution. \edit{Latitudinal variations of the surface temperature, as on Mars (\cite{kieffer2013}) do not have a significant impact on the long-term evolution and large-scale dynamics of the planet (\cite{plesa2016}) and thus can be safely neglected.} The temperature of core-mantle boundary evolves as the core cools (eq. \ref{eq-CoreCool}). There is no heat flux across the side walls of the computational domain, i.e. they are assumed to be insulating (see Fig. \ref{fig-sketch}). Table \ref{table-Mars-paras-1} and Table \ref{table-Mars-paras-2} list all the fixed dimensional parameters that are shared by all simulations. 

We ran several single-core simulations using the finite-volume code GAIA \cite{huettig2013} with a grid resolution of $300$ radial layers and $392$ cells per layer. Even though $3251$ out of $10,525$ simulations did not reach the end time of $4.5$ Gyr (i.e. time since formation of the planet until today), we use all the time-steps available from all the simulations. \edit{Some combinations of parameters can make the system of PDEs too stiff to efficiently solve, as the time-stepping becomes increasingly small. This means that some simulations can take more than $50$ times the average run time.} In total, the dataset amounts to $10$ TB and took approximately $2$ million CPU hours. \edit{The challenging task of predicting the 2D temperature field for a variety of parameters necessitated this computational effort. The number of simulations used in this study is therefore almost a $100$ times that of what a typical parameter study in mantle convection would use.} The distribution of the parameters for all the simulations in the training, cross-validation and the test set is plotted in Fig. \ref{fig-paras}. 

\section{Compression of temperature fields}
\label{sec-compress}

\begin{figure*}
\centering
\includegraphics[width=\textwidth]{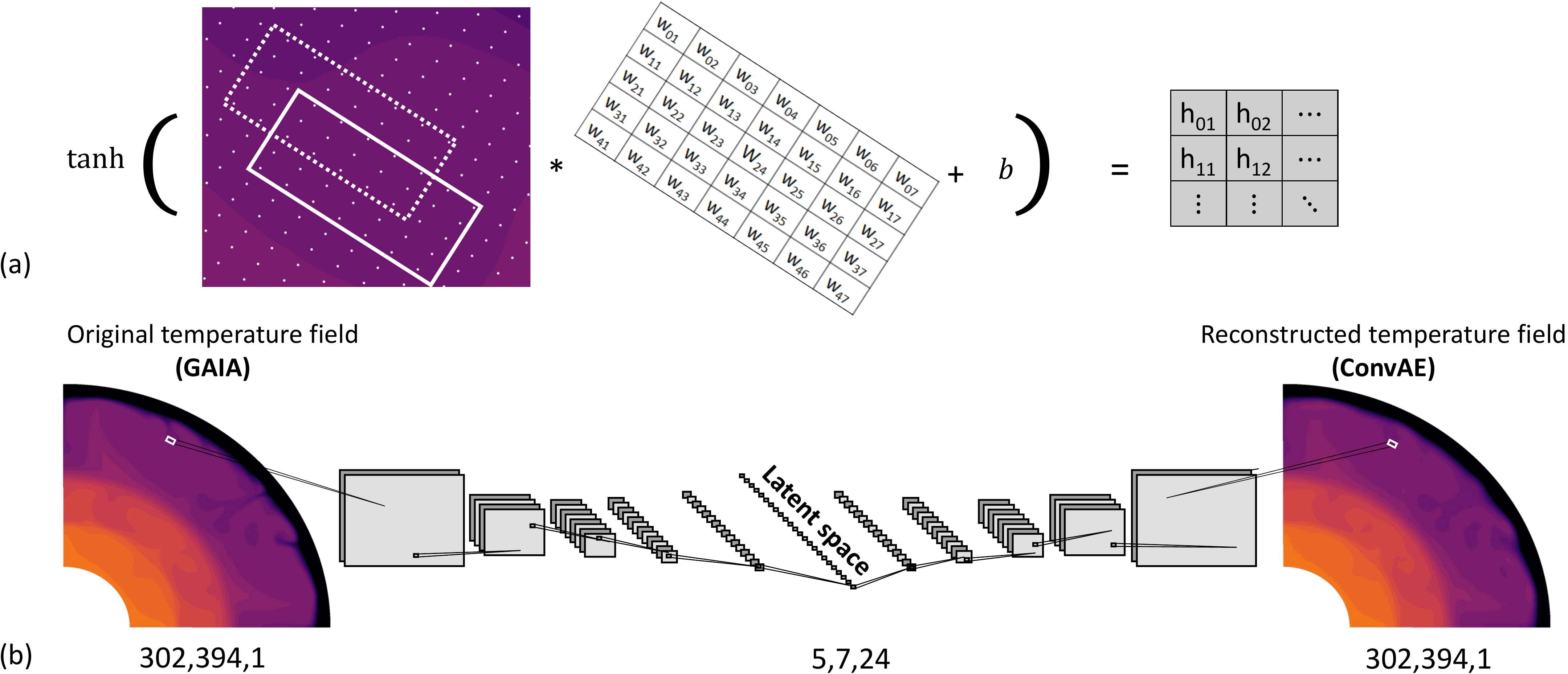}
\caption{Dimensionality reduction of the 2D temperature fields using convolutional autoencoders. (a) Filters with trainable parameters ($\mathbf{w}$) move across the computational domain with specified strides. After the convolution operation ($\ast$), the bias $b$ is added to it before applying the activation function, resulting in the entries for the next hidden layer $\mathbf{h}$. (b) Several filters can be used to successively reduce the size of the original field $(302 \times 394 \times 1)$ till a desired encoding or latent space representation is obtained $(5 \times 7 \times 24)$. The convolutional autoencoder then reconstructs the compressed field back to its original dimensions $(302 \times 394 \times 1)$ using deconvolution operations and optimizes the weights using the difference between the prediction and the ground truth. }
\label{fig-convae}
\end{figure*}

The temperature fields alone account for approximately $1$ TB. While most everyday computers cannot hold that much data in memory, one could still overcome the issue using a ``data-generator''. A data-generator feeds the ML algorithm during training and/or inference by reading and storing only a limited number of examples from the disc at any given time. Nevertheless, these require careful programming to ensure that the CPU or GPU are constantly fed batches of data without having to wait for the next one. Furthermore, performing ML calculations on the full-sized fields would be slower than a compressed version of it and introduce a lot more trainable parameters which could increase the risk of over-parameterization. Hence, we decided to compress the temperature fields first and then scan for different ML algorithms and architectures that can help us predict this latent space representation. 

Traditionally, linear reduced-order modelling techniques like proper orthogonal decomposition (POD) have been used for truncating high-fidelity simulations using the most dominant modes \cite{lumley_1967}. However, the orthonormal bases (typically obtained through singular value decomposition) of one simulation do not generalize well onto those of another and often require non-trivial basis interpolation (e.g., \cite{Friderikos2020}). Recently, \cite{Mohan2020} demonstrated that convolutional autoencoders (ConvAE) \cite{convae} provide a powerful non-linear tool for compressing flow fields, bypassing the need for calculating POD modes. 

Instead of fully connected dense layers like conventional autoencoders, ConvAE uses convolutional filters. As shown in Fig. \ref{fig-convae}(a), a filter with trainable weights $\mathbf{w}$ moves across the state variable field (temperature) as specified by a hyperparameter called ``stride''. A stride of $2$, for example, means that the filters move two units (two numerical grid cells) horizontally and then when a row is completed, two units vertically. Fig. \ref{fig-convae}(a) shows a filter with height $5$ and length $7$ (also hyperparameters) convolving with the temperature field at strides of $2$ in both $x-$ and $y-$direction. We use $\tanh()$ as activation function, which, when applied to the sum of the bias and the convolution product, introduces non-linearity and returns the output for the next hidden layer which can then be convolved on. This process continues until the desired latent space representation is reached (Fig. \ref{fig-convae}(b)). Then, the compressed state can be successively restored to the original size with the help of another sequence of convolution filters, called deconvolutions in this context. Once the forward graph has been setup, one can then minimize the difference between the original and reconstructed 2D temperature field by back-propagating the derivative of the error with respect to the network weights. \cite{convae} point out that a convolutional architecture offers two main benefits over the fully connected dense layers: (1) the 2D structures of the convolutional filters retain spatial correlations which would otherwise be lost in 1D dense layers and (2) ConvAE have significantly fewer trainable parameters due to the shared weights.

We use Keras, an API built on top of Tensorflow \cite{chollet2015keras} for training our ConvAE. We split the simulations into $98\%$, $1\%$ and $1\%$ for training, validating and testing, respectively. Given the size of the entire training set, we only feed the GPU mini-batches of $16$ temperature fields (i.e. time-steps of any simulation) during training. We use L2 regularization and early-stopping by manually monitoring the validation loss (mean-squared error) and the optimization is carried out using Adam \cite{adam}. 

\begin{figure*}
\centering
\includegraphics[width=0.7\textwidth]{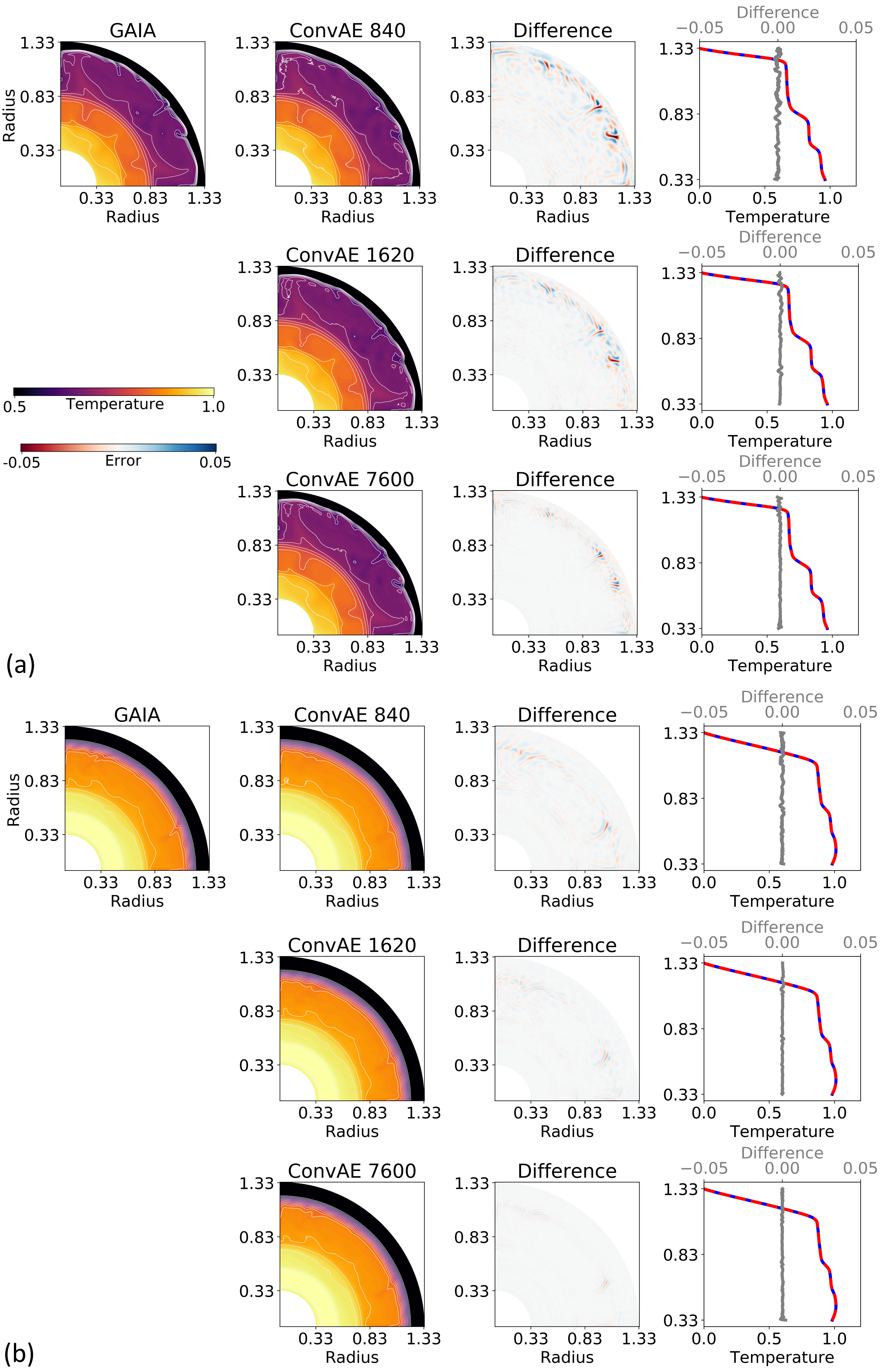}
\caption{(a) First example of compression and reconstruction from the test set for different architectures of convolutional autoencoders. ConvAE $840$ represents a $142$-fold compression of the original field ($302 \times 394$), while ConvAE $1620$ compresses the field by a factor of $73$ and ConvAE $7600$ by a factor of $16$. (b) Second example with vigorous convection and small-scale structures from the test set. For each ConvAE, the error in the reconstruction is plotted in the third column, along with the horizontally-averaged 1D temperature profiles in the fourth column (and the difference between the two).}
\label{fig-results-convae}
\end{figure*}

Since, the \edit{computer} cannot hold the entire training-set in memory, we use a data-generator. Using multi-processing built into Keras' ``fit-generator'' for creating multiple batches in parallel along with multi-threading for populating each batch with the help of Joblib \cite{joblib}, it takes around $6$ hours for one epoch to complete. After $5$--$10$ epochs, acceptable results are obtained. Fig. \ref{fig-results-convae}(a) and \ref{fig-results-convae}(b) show results of reconstruction for two different examples in the test set using three different ConvAE architectures, which differ in the dimensions of the latent space. As it can be expected, the more the temperature fields are compressed, the less accurate the reconstruction is. In all the plots of the temperature field in this paper, we clip the colorbar below 0.5 and above 1.0 to enhance the visualization of plumes and downwellings. Furthermore, we always plot the non-dimensionalized temperature and the non-dimensionalized radius. 

We find that ConvAE with a latent size of $840$ (or width $\times$ height $\times$ channels $=$ $5 \times 7 \times 24$) offers an excellent compression factor of $142$, while being able to reconstruct the temperature fields with a mean relative accuracy of $99.80\%$ on the test set. In comparison, ConvAEs with $1620$- and $7600$-dimensional latent spaces are $99.88\%$ and $99.90\%$ accurate, respectively. To calculate the \edit{mean relative accuracy}, we dimensionalize the temperature using Eq. \eqref{eq-EBA-ND-4} to avoid division by zeros. 

In the subsequent sections, we discuss the training of ML algorithms to predict the temperature fields compressed to $840$ numbers and then reconstruct the $302\times394$-sized temperature fields using the trained decoder, i.e. we can compress the data from $1$ TB to approximately $7$ GB. \edit{We also tested the $1620$- and $7600$-dimensional encodings when predicting the compressed temperature fields - as explained in the subsequent section - and found no significant improvement in the predictions, suggesting that the accuracy of the $840$-dimensional compression is good enough for a dataset of this size. Keeping the latent space as small as possible is especially useful when training LSTMs, because an LSTM cell has $8$ times as many trainable parameters as a dense FNN layer.}

The results of the simple ConvAE are also encouraging in the light that the computational grid is structured, but not uniform, as is typically the case in computer vision applications. It seems that the different filters in the ConvAE are capable of capturing features at different spatial scales. For now, accounting for the curvilinear nature of the mesh and thereby potentially achieving higher compressibility and/or accuracy remains subject to future research.  

\section{Prediction of compressed temperature fields}
\label{sec-predict}
\subsection{Neural networks}
\label{sec-nn}

With the temperature fields compressed, we now move on to predicting this latent space representation of $840$ numbers, based on the five parameters of the simulation. Since \cite{agarwal2020} demonstrated that an FNN which takes the five parameters as inputs and time as a sixth input is capable of predicting the 1D temperature profile with a high accuracy, FNN is an obvious candidate for predicting the compressed temperature fields. 

\begin{figure*}
\centering
\includegraphics[width=\textwidth]{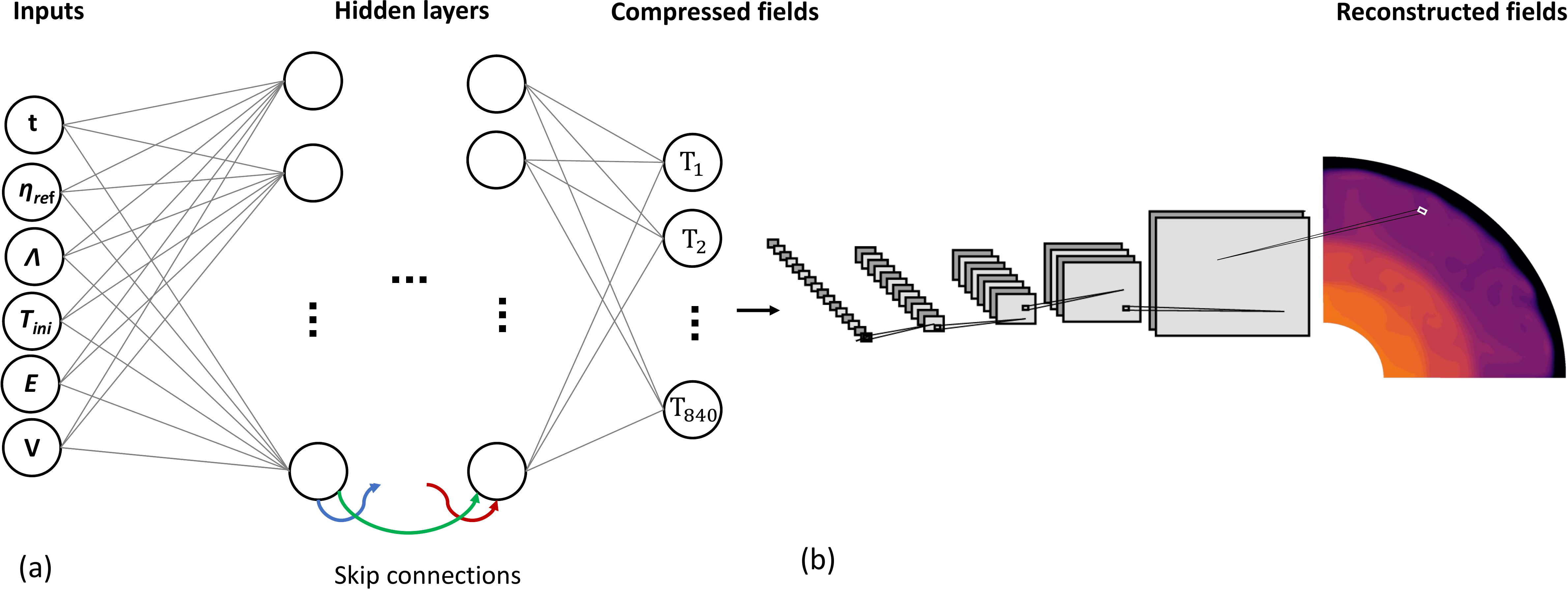}
\caption{(a) The five parameters governing mantle convection and time are used to predict the compressed temperature fields. Skip connections are used to add the output of each hidden layer after activation to that of each of the following hidden layers before activation. (b) After the training is complete, the trained decoder from ConvAE is used to reconstruct the temperature field back to its original dimensions.}
\label{fig-nn}
\end{figure*}

\begin{figure}
\centering
\includegraphics[width=0.3\textwidth]{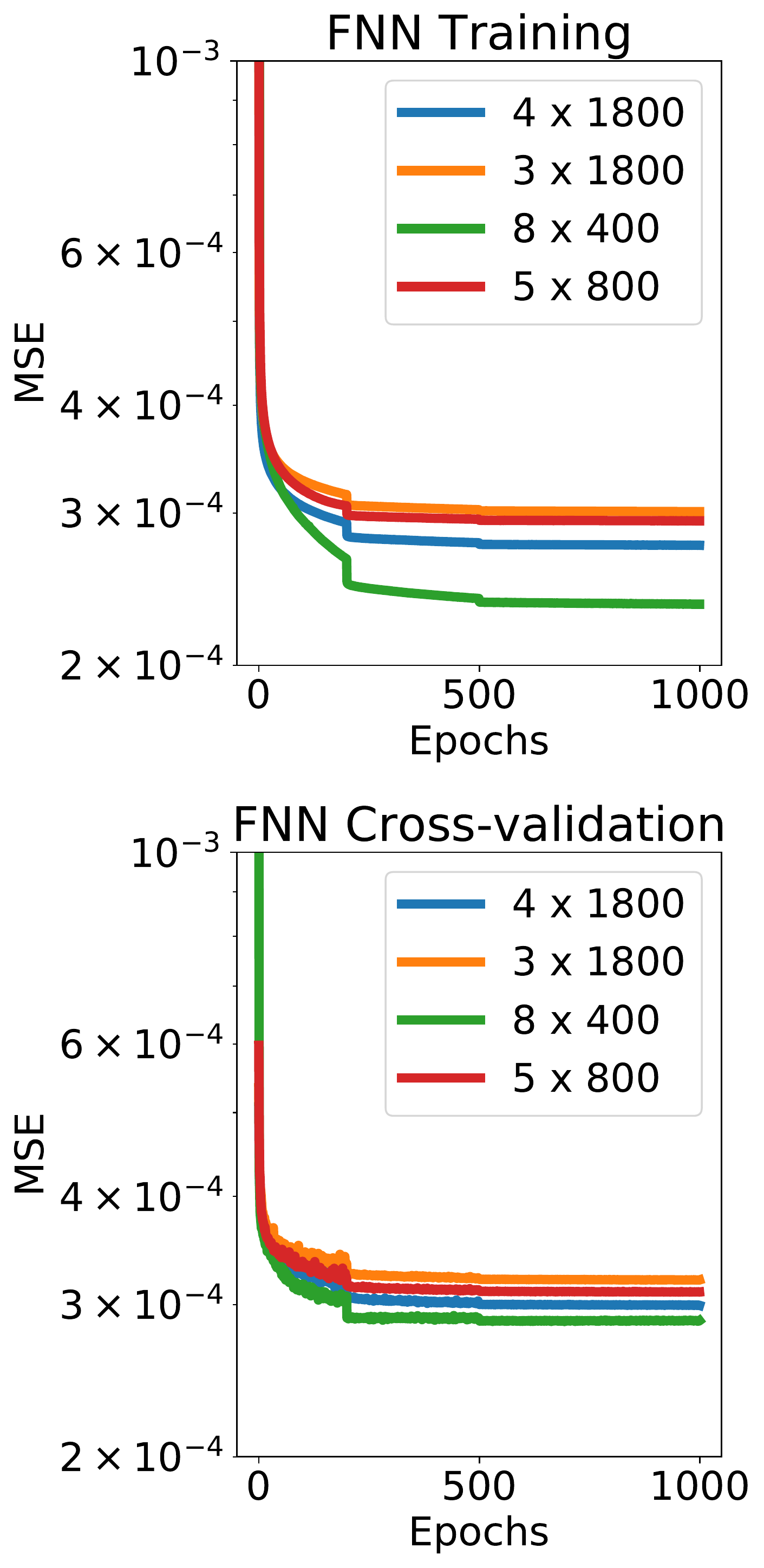}
\caption{The evolution of the mean-squared error (MSE) on (a) training data and on (b) cross-validation data for different FNN architectures. The legend shows the number of hidden layers as well as the number of neurons per layer of a given FNN architecture. For example, $4 \times 1800$ means the network has four hidden layers with $1800$ neurons each. \edit{The step-like drop after $200$ epochs is a result of the decrease in the learning rate.}}
\label{fig-nn-loss}
\end{figure}

\begin{figure*}
\centering
\includegraphics[width=0.88\textwidth]{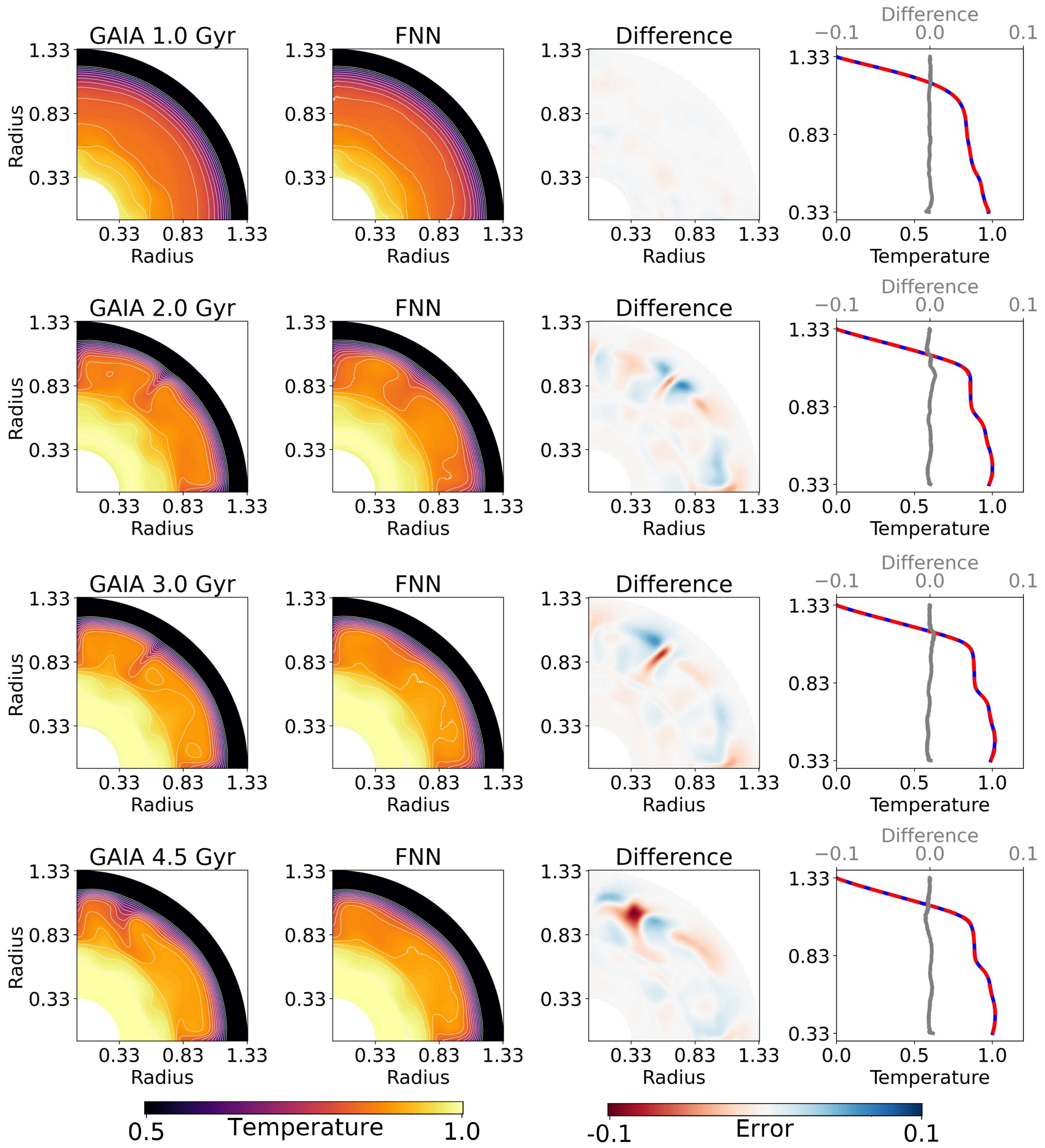}
\caption{Example of a sluggishly convecting mantle from the test set. The parameters are: $\eta_{\rm ref}= 3.6 \times 10^{21}$ Pa S, $E = 1.6 \times 10^{5}$ J mol$^{-1}$, $V = 4.4 \times 10^{-6}$ m$^3$ mol$^{-1}$, $\Lambda = 15.3$ and $T_{\rm ini} = 1634$ K. The temperature field from GAIA and its equivalent FNN prediction are shown in column 1 and 2, respectively. The third column shows the difference between the two. Column 4 shows the horizontally-averaged 1D temperature profiles from GAIA ( solid blue) and FNN (dashed red), as well as the difference between the two (grey).}
\label{fig-nn1}
\end{figure*}

\begin{figure*}
\centering
\includegraphics[width=0.88\textwidth]{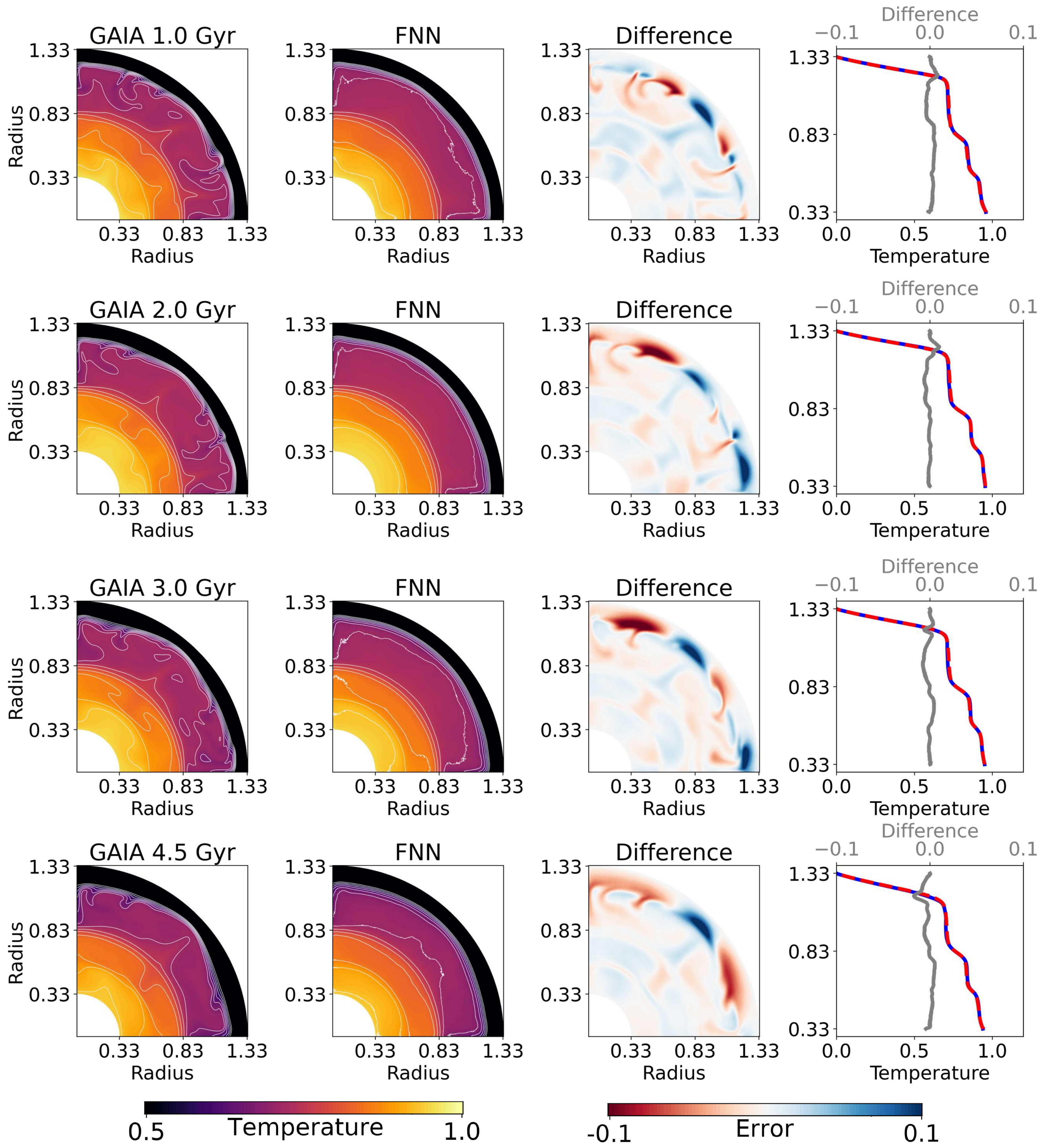}
\caption{Example of a vigorously convecting mantle from the test set. The parameters are: $\eta_{\rm ref}= 5.0 \times 10^{19}$ Pa S, $E = 1.5 \times 10^{5}$ J mol$^{-1}$, $V = 7.6 \times 10^{-6}$ m$^3$ mol$^{-1}$, $\Lambda = 30.7$ and $T_{\rm ini} = 1705$ K. The temperature field from GAIA and its equivalent FNN prediction are shown in column 1 and 2, respectively. The third column shows the difference between the two. Column 4 shows the horizontally-averaged 1D temperature profiles from GAIA ( solid blue) and FNN (dashed red), as well as the difference between the two (grey).}
\label{fig-nn3}
\end{figure*}

Fig. \ref{fig-nn}(a) shows the five parameters and the time that are used as inputs to the FNN to predict the compressed temperature fields. For computational efficiency, we optimize the network weights using small mini-batches of $16$ temperature fields and Adam. Also, we used Scaled Exponential Linear Unit (SELU) as our activation function \cite{selu}:
\begin{equation}
\label{eq-selu}
    SELU(x) = \lambda \left \{ \begin{matrix}
x\\ 
\alpha e^{x} - \alpha
\end{matrix} \right.,
\end{equation}
as it seemed to deliver slightly better results than $\tanh$. In Eq. \eqref{eq-selu}, $\alpha=1.67326324$ and $\lambda=1.05070098$ are pre-defined \edit{and come from the original paper \cite{selu}}. We further schedule the learning rate to decrease by a factor of $10$ after $200$ epochs and then again by a factor of $10$ after the next $300$ epochs. During the training, we only save the network weights if the validation loss drops. Furthermore, we employ a dropout of $5\%$ after each hidden layer as a further regularization. Once the training has finished, the 2D fields are reconstructed from the predicted latent states as a post-processing step using the already trained decoder. 

We trained different FNN architectures with fully connected dense layers. The evolution of the mean squared error (MSE) on the training and the cross-validation data is plotted in Fig. \ref{fig-nn-loss}. The cross-validation loss over epochs converges to approximately the same value, especially for deeper networks. Since, we also test some deep architectures such as five hidden layers with $800$ neurons each and eight hidden layers with $400$ neurons each, we use skip connections from each hidden layer to all the following hidden layers via addition. Given the challenging task of predicting a $840$--dimensional vector, we use fairly deep architectures. Therefore, we employ SELU activation and skip connections to overcome the problem of vanishing gradients typically observed in deep networks.

For the FNN with $8$ hidden layers of $400$ units each, we take two examples from the test set with different convection patterns and plot one characterized by a sluggish behavior in Fig. \ref{fig-nn1} and another one characterized by vigorous convection in Fig. \ref{fig-nn3}. \edit{A schematic of the this particular FNN architecture is provided in Appendix \ref{Section_Appendix_FNN}}. The figures show the 2D temperature fields computed with GAIA (first column), those predicted by the FNN (second column), the difference between the two (third column), as well as the horizontally-averaged 1D temperature profiles along with their difference (fourth column). While the 1D profiles show good agreement (as previously demonstrated by \cite{agarwal2020}), the 2D predictions show more inaccuracies. In particular, cold, sub-lithospheric downwellings, which are a fundamental feature of the planform of stagnant-lid convection, tend to be completely lost by the FNN prediction. See Supplemental Material at \cite{supplementary_files} for animations of these two examples along with three further examples from the test set. In comparison with the GAIA simulations, the FNN predictions also fail to capture the vigor of convection. Even when the FNN captures a downwelling early on in the evolution, its lateral transport is not captured. On average, the 2D temperature fields predicted by FNN are $99.30\%$ accurate (\edit{mean relative accuracy} of dimensionalized temperature fields) with respect to GAIA and $99.35\%$ with respect to ConvAE. 

\subsection{Long short-term memory (LSTM)}

\begin{figure}
\centering
\includegraphics[width=0.43\textwidth]{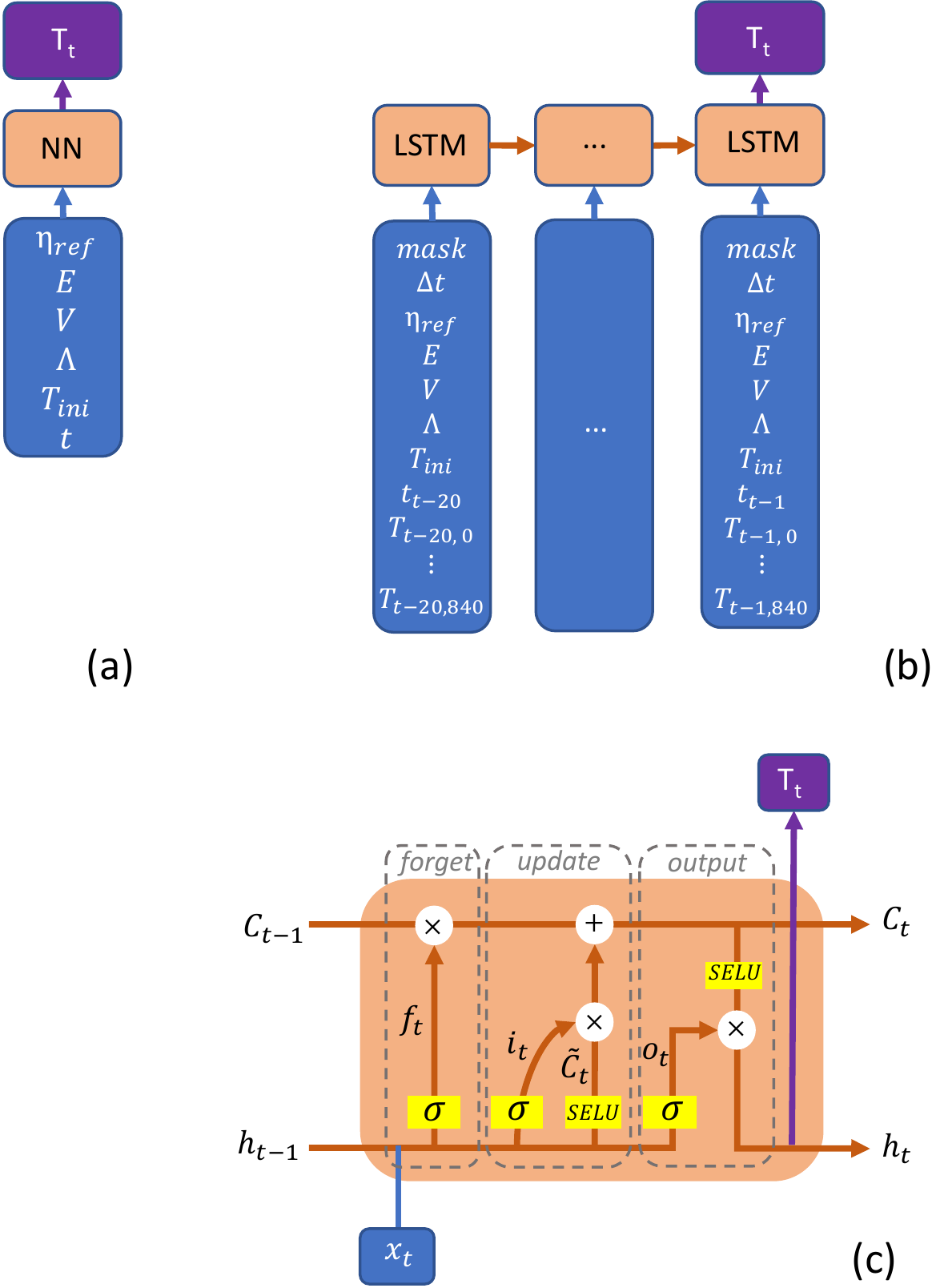}
\caption{Comparison of how (a) the FNNs are trained and how the (b) many-to-one LSTMs are trained for predicting compressed the temperature field $T_t$. (c) Each LSTM cell receives as input the compressed temperature field at time $t-1$ along with other parameters: a mask for whether the next time-step exists, difference between the time-step used as input and the one being predicted ($\Delta t$), the five mantle convection parameters ($\eta_{\rm ref}$, $E$, $V$, $\Lambda$ and $T_{\rm ini}$), as well as the time itself $t_{t-1}$ of the input compressed temperature field. \edit{In practice, the mask is used by the data-generator to decide whether to provide the next time-step as output to train the LSTM on.}}
\label{fig-lstm}
\end{figure}

The failure of the FNNs to capture lateral convection structures such as down- and upwellings can be attributed in large part to the fact that the temporal snapshots of any given simulation are disconnected. By treating time as an additional input variable and shuffling time-steps of different simulations (but within the training/validation/test splits), the details of the dynamics of the flow are largely lost. 

Hence, we turn our attention to recurrent neural networks that have been shown to be successful for a variety of Natural Language Processing tasks. Recurrent architectures such as LSTM \cite{lstm} provide a back-propagation mechanism acting through a sequence of inputs (such as time-steps of a simulation), thereby allowing the network to learn temporal dynamics (e.g., \cite{Mohan2020, Hamidreza2020}).

As shown in Fig. \ref{fig-lstm}(b), LSTMs differ from an FNN architecture (Fig. \ref{fig-lstm}(a)) in that they use a set of previous time-steps to predict the next time-step. Furthermore, each LSTM cell (Fig. \ref{fig-lstm}(c)) is more elaborate than a simple neuron of a fully-connected FNN. \cite{olah_2015} provides an accessible explanation of LSTMs. A brief description of an LSTM cell is provided in Sec. \ref{Section_Appendix_LSTM}.

As before, we use Keras API for setting up the forward graph (Eqs. \eqref{eq-lstm-a}--\eqref{eq-lstm-e}) and optimizing the trainable parameters by minimizing the MSE between the prediction and the target. Since, the time-steps for simulations were stored after a specified number of time-steps of the numerical solver, as well as at every physical time-interval of $100$ million years of planetary evolution, we ended up with non-uniform time-series. Hence, we use ``masking'', as done for example by \cite{missingRNN}, to specify if there is a time-step to predict ($mask=1$) or not ($mask=0$). Most of the simulations have less than $200$ time-steps, although $9$ simulations exceed $400$ time-steps. 

In addition to the mask, we also input the size of the time-step ($\Delta t$), the five model parameters, time and the compressed temperature field at the current time ($t$). After some trial and error, we found $20$ time-steps to serve as a rich-enough input to predict the $21$-st. When predicting a time-step when $20$ previous inputs are not available, say for time-step $10$, we took time-steps $0$ through $9$ and filled the rest with time-step $0$. This way, the thermal evolution of a planet can be simulated based purely on input parameters. The input shape of each mini-batch is, thus, (simulations$=16$, time-steps$=20$, input$=848$) and the output shape is  (simulations$=16$, time-steps$=1$, output$=840$).

\begin{figure}
\centering
\includegraphics[width=0.3\textwidth]{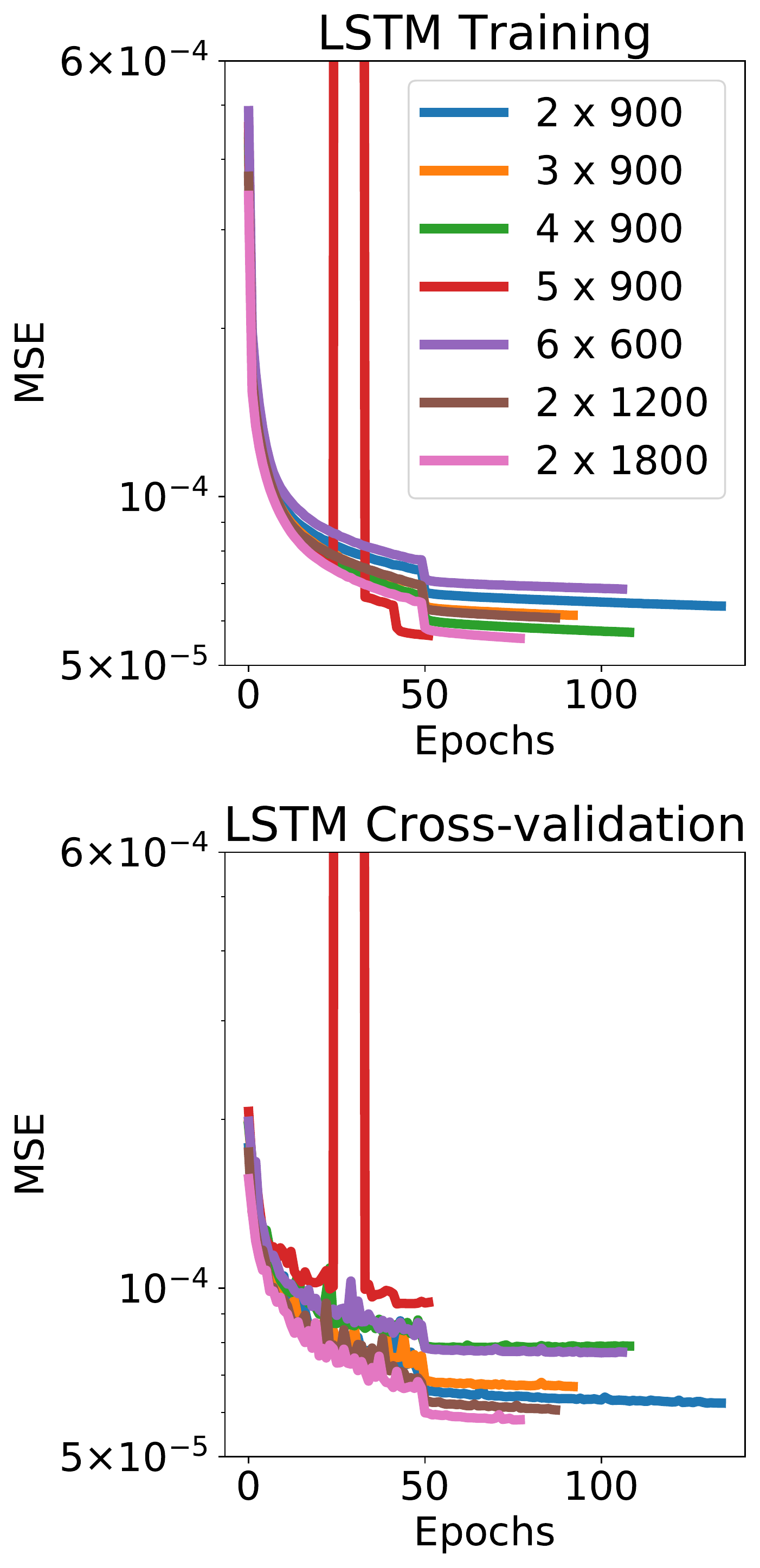}
\caption{The evolution of MSE on (top panel) training data and (bottom panel) cross-validation data for different LSTM architectures. The legend shows the number of hidden layers as well as the number of LSTM cells per layer. For example, $4 \times 900$ means the network has four hidden layers with $900$ cells each. \edit{The step-like drop after $50$ epochs is a result of the decrease in the learning rate.}}
\label{fig-lstm-loss}
\end{figure}

The four different gates with two sets of matrices each in an LSTM cell mean that there are $8$ times as many trainable parameters per hidden layer as a regular dense layer in an FNN. Therefore, we tested a limited number of LSTM architectures as shown in Fig. \ref{fig-lstm-loss}. For computational efficiency, we use $SELU$ again, instead of $tanh$ as activation function \cite{selu_lstm}. 
Such models take about $2$ weeks on a Tesla V100 GPU to reach asymptotic loss values. Just as in the case of training FNNs, we use the following strategies to prevent over-fitting: (1) Storing the weights only if the validation loss drops, (2) using a dropout of $5\%$ for each hidden layer, and (3) training on mini-batches of $16$. 

The difference between the loss curves for different architectures is small and given the stochasticity associated with training of LSTMs, not particularly enlightening with one small exception. The loss curve for the LSTM with five hidden layers of $900$ cells each diverged around epoch 20, but seems to have found its way back a few epochs later, indicating that such a deep architecture with roughly $33$ million trainable parameters might be prone to exploding gradients, especially when the learning rate is too large (0.0001). Of course, the initial learning rate is divided by a factor of $10$ after the first $50$ epochs and then by another factor of $10$ after the next $150$ epochs (if reached).

\begin{figure*}
\centering
\includegraphics[width=0.7\textwidth]{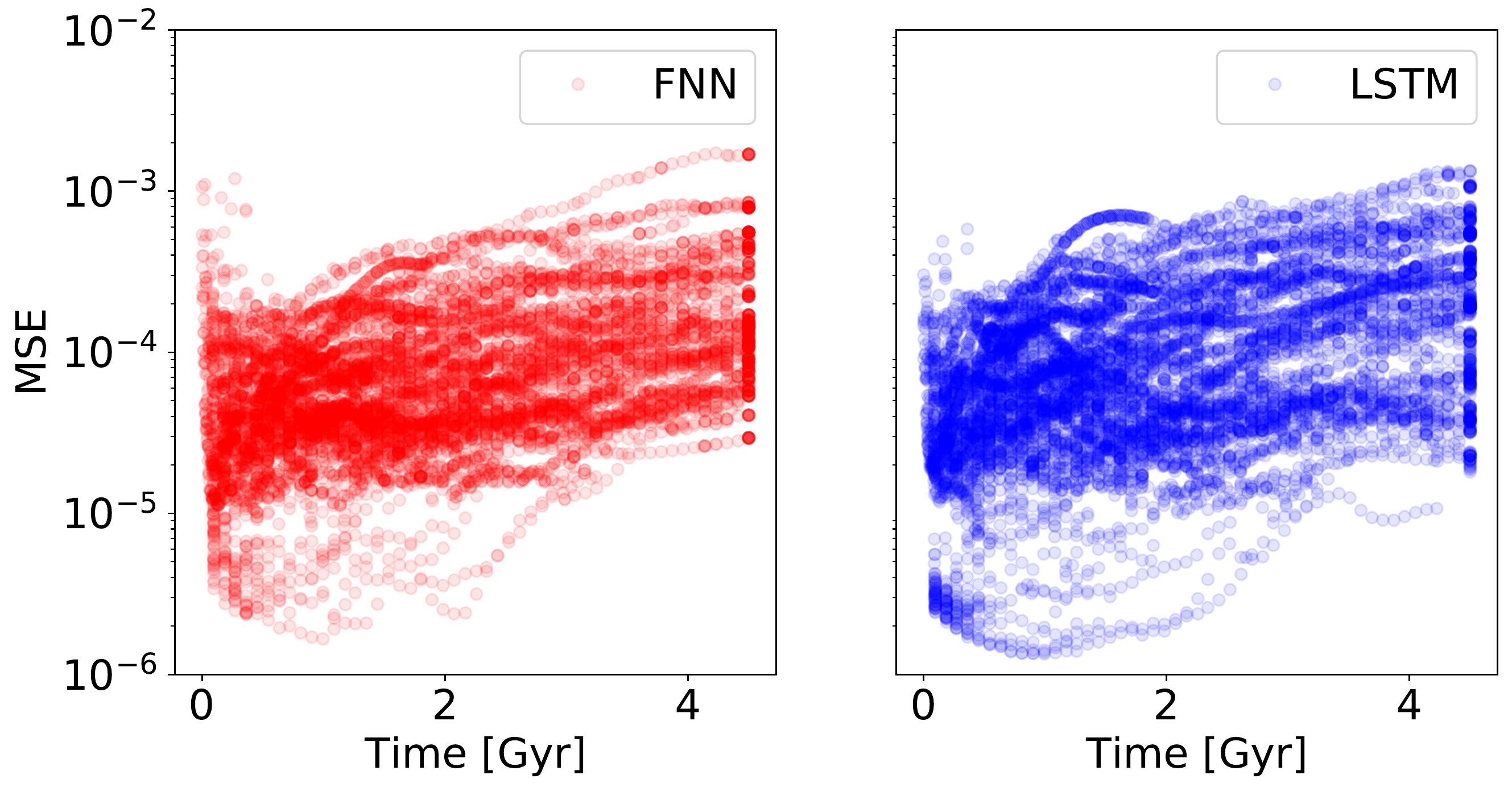}
\caption{\edit{Spatially averaged MSE for FNN ($8 \times 400$) and LSTM ($4 \times 600$) vs. physical time for each simulation in the test set.}}
\label{fig-timeError}
\end{figure*}

LSTM architectures achieve a lower MSE loss (Fig. \ref{fig-lstm-loss}) than FNN architectures (Fig. \ref{fig-nn-loss}). However, it is important to note that the MSE loss displayed for LSTMs during training is based on a prediction, which takes the highly accurate compressed temperature fields as inputs. In inference mode, as the network iteratively predicts the next time-step based on the previously predicted 20 time-steps, this accuracy might decrease due to accumulation of error. Therefore, we calculated the \edit{mean relative accuracy} for all the simulations in the cross-validation set using different LSTMs in purely inference mode and display them along with the \edit{mean relative accuracy} on the test set in Table \ref{table-lstm-accuracy}. 
\setlength{\tabcolsep}{1em}
\begin{table}
\centering
\caption{\edit{mean relative accuracy} of different LSTM architectures on the cross-validation (CV) and the test sets, when computed in inference mode (i.e. each input temperature field is also an LSTM prediction). For reference, \edit{mean relative accuracy} for the FNN architecture is also presented.}
\label{table-lstm-accuracy}
\begin{tabular}{ccc}
Accuracy & \textbf{wrt GAIA} $(\%)$ & \textbf{wrt ConvAE} $(\%)$ \\
\hline 
\hline
$\mathbf{2}$ $\mathbf{\times}$ $\mathbf{900}$ & &  \\
Test &  98.182 $\pm$ 9.853   &  98.237 $\pm$ 9.854 \\
CV   &  96.542 $\pm$ 15.041  &  96.597 $\pm$ 15.049 \\
\hline
$\mathbf{3}$ $\mathbf{\times}$ $\mathbf{900}$ & &  \\
Test &  99.222 $\pm$ 0.515 &  99.278 $\pm$ 0.513 \\
CV   &  99.109 $\pm$ 0.602 &  99.164 $\pm$ 0.605 \\
\hline
$\mathbf{4}$ $\mathbf{\times}$ $\mathbf{900}$ & &  \\
Test &  99.226 $\pm$ 0.524 & 99.285 $\pm$ 0.525 \\
CV   &  99.082 $\pm$ 0.674 & 99.141 $\pm$ 0.682 \\
\hline
$\mathbf{5}$ $\mathbf{\times}$ $\mathbf{900}$ & &  \\
Test &  99.199 $\pm$ 0.537 & 99.257 $\pm$ 0.537 \\
CV   &  99.051 $\pm$ 0.680 & 99.108 $\pm$ 0.687 \\
\hline
$\mathbf{4}$ $\mathbf{\times}$ $\mathbf{600}$ & &  \\
Test &  99.221 $\pm$ 0.495 & 99.281 $\pm$ 0.495 \\
CV   &  99.081 $\pm$ 0.665 & 99.140 $\pm$ 0.671 \\
\hline
$\mathbf{2}$ $\mathbf{\times}$ $\mathbf{1200}$ & &  \\
Test &  98.275 $\pm$ 8.313  & 98.328 $\pm$ 8.317 \\
CV   &  97.364 $\pm$ 11.352 & 97.416 $\pm$ 11.356 \\
\hline
$\mathbf{2}$ $\mathbf{\times}$ $\mathbf{1800}$ & &  \\
Test &  98.561 $\pm$ 6.446 & 98.616 $\pm$ 6.445 \\
CV   &  98.326 $\pm$ 7.376 & 98.379 $\pm$ 7.380 \\
\hline
\hline
\textbf{FNN} & &  \\
Test &  99.297 $\pm$ 0.433 & 99.354 $\pm$ 0.422  \\
CV   &  99.207 $\pm$ 0.482 & 99.262 $\pm$ 0.475 \\
\end{tabular}
\end{table}
Table \ref{table-lstm-accuracy} seems to contradict the results of the MSE plots in Fig. \ref{fig-lstm-loss}. While during training, the $[1800, 1800]$ LSTM with $46.5$ million trainable parameters attained the lowest MSE, in inference mode it had the third lowest \edit{mean relative accuracy} of all architectures. In contrast, the $[600, 600, 600, 600]$ LSTM with $12.7$ million trainable parameters, achieved a higher accuracy on the cross-validation set and on the test set. We simply do not have enough simulations to fit $46.5$ million weights without over-fitting, which is evident from the high standard deviation of absolute relative accuracy on test and cross-validation sets. In inference mode, the errors in predicting time-steps can accumulate to the point where a simulation diverges. Luckily, this behavior is not observed in the $[600, 600, 600, 600]$ LSTM, for example. We present the evolution of the averaged MSE with time for different simulations in the test set, for both the FNN and LSTM, in Fig. \ref{fig-timeError}. \edit{As the mantle typically starts cooling after some point in the thermal evolution, the convection should get slightly less vigorous. This means that the upwellings and downwellings should have a longer wavelength and therefore, become slightly easier to predict. Nevertheless, it seems that the lack of data towards the end of the evolution (from unfinished simulations) is the reason for the increased error. In case of LSTMs it can be further exacerbated by the accumulation of error.}

For both the LSTM and the FNN, the error tends to increase marginally with time. This can be attributed in large part to the fact that data get sparser with increasing time as not all simulations finished. For the LSTM though, the accumulation of error \edit{can be} another contributing factor. 

\begin{figure*}
\centering
\includegraphics[width=0.9\textwidth]{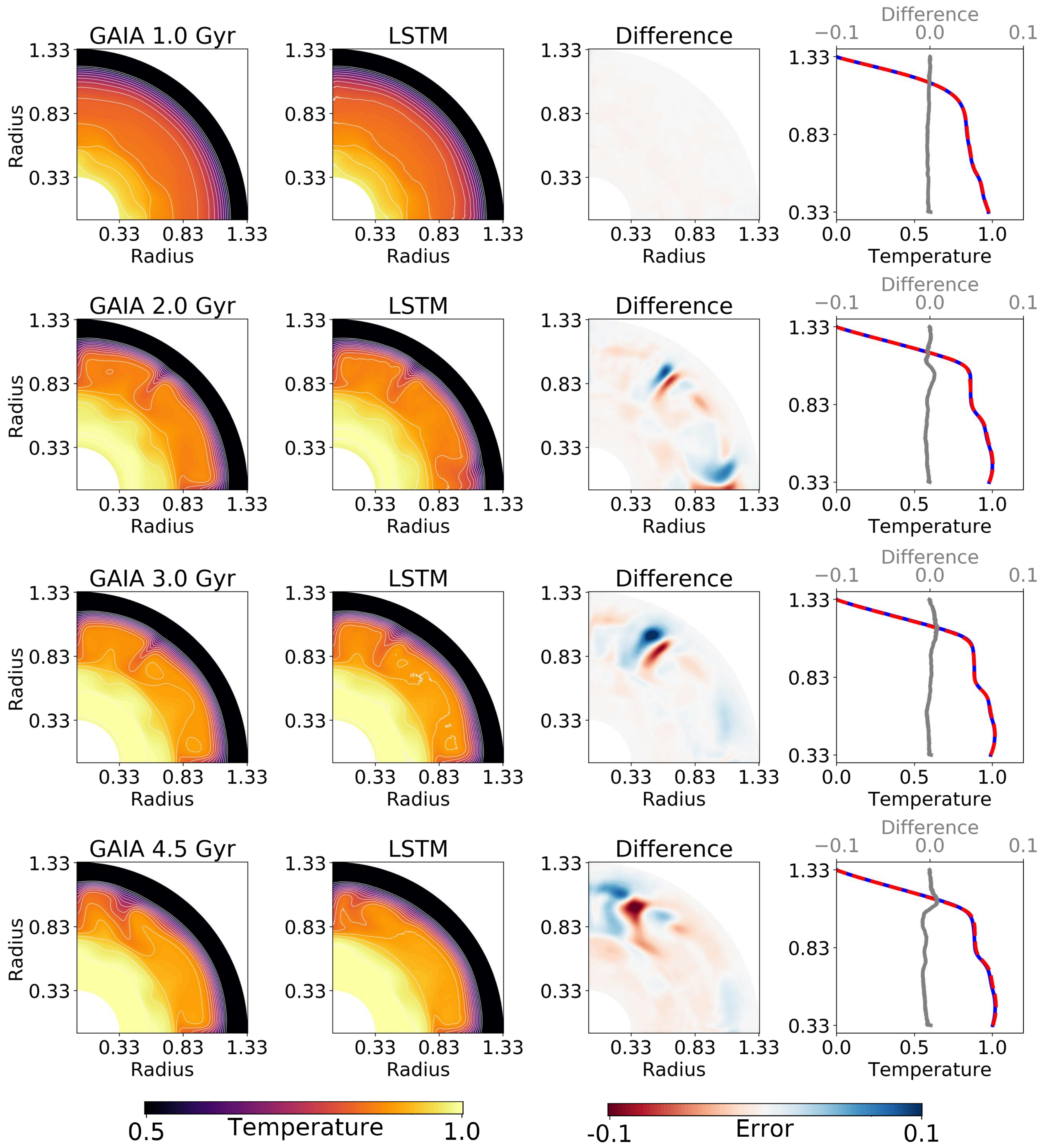}
\caption{Example 1 from the test set. The temperature field from GAIA and its equivalent LSTM prediction are shown in column 1 and 2, respectively. The third column shows the difference between the two. Column 4 shows the horizontally-averaged 1D temperature profiles from GAIA (solid blue) and LSTM (dashed red), as well as the difference between the two (grey).}
\label{fig-lstm1}
\end{figure*}

\begin{figure*}
\centering
\includegraphics[width=0.9\textwidth]{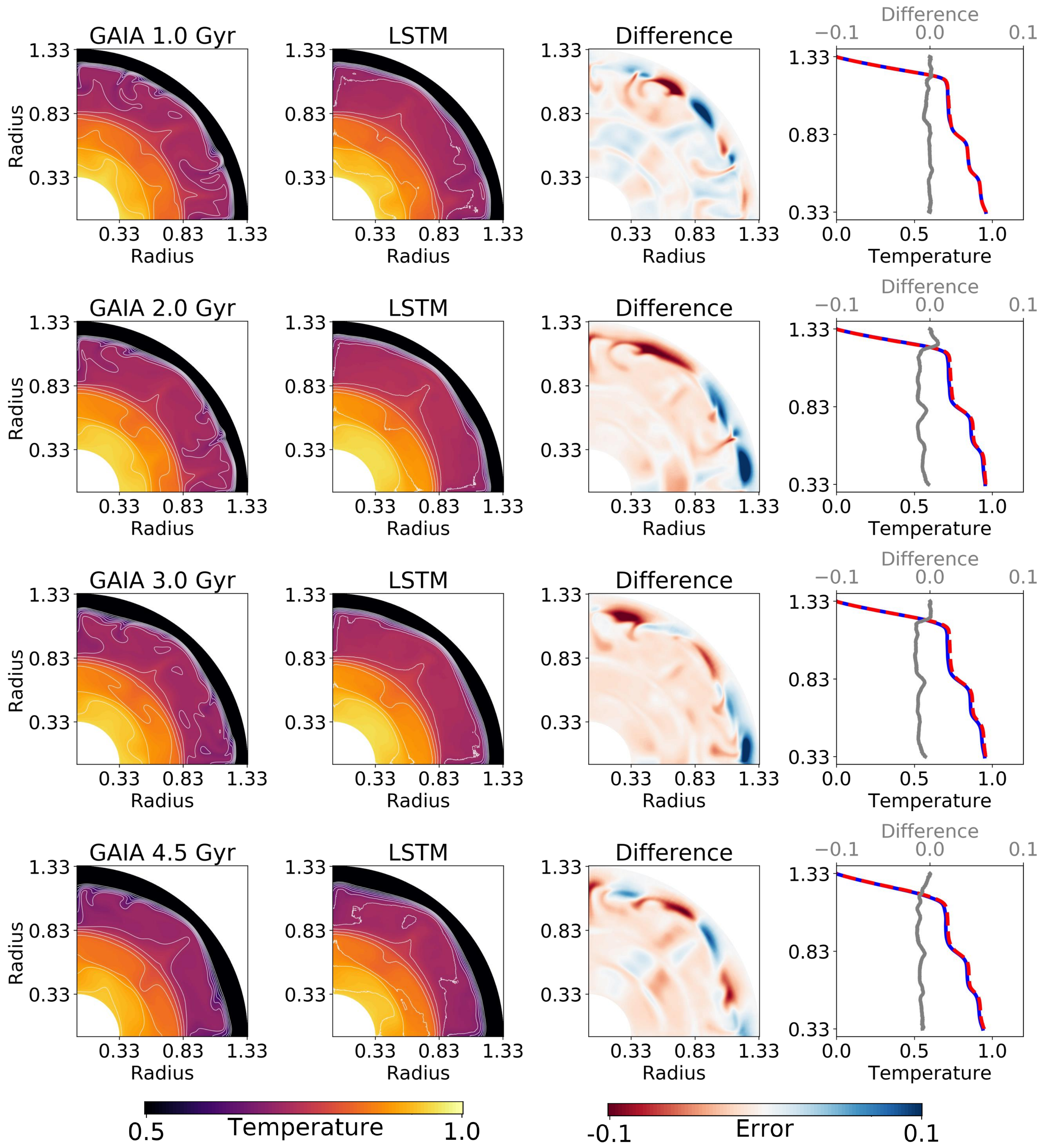}
\caption{Example 2 from the test set. The temperature field from GAIA and its equivalent LSTM prediction are shown in column 1 and 2, respectively. The third column shows the difference between the two. Column 4 shows the horizontally-averaged 1D temperature profiles from GAIA (solid blue) and LSTM (dashed red), as well as the difference between the two (grey).}
\label{fig-lstm3}
\end{figure*}

In Fig. \ref{fig-lstm1} and Fig. \ref{fig-lstm3}, we plot the same two examples from the test set as in subsection \ref{sec-nn}, but using the $[600, 600, 600, 600]$ LSTM this time. 
Despite the slightly lower \edit{mean relative accuracy} compared to FNN, the LSTMs do a better job of capturing the convection structures. This is especially true for the more sluggish simulations such as the one in Fig. \ref{fig-lstm1}, where the downwelling is not only formed at $2$ Gyr (second row)(), but also maintained and transported towards the left boundary in time, unlike the FNN.  

Fig. \ref{fig-lstm3} is a good example of how the small-scale downwellings formed under more vigorous convection are not well captured by the LSTM. This explains the richness of structures one can see from the difference plots in column 3. Nevertheless, the big downwelling captured at 1 Gyr to the right of the domain at radius of $0.83$ to $1.1$, for example, or the upwelling at same radial location, but towards the middle of the domain present an improvement over the smudged-out prediction of the same simulation by an FNN in Fig. \ref{fig-nn3}. 

In summary, LSTMs are better at predicting sharper structures such as downwellings as well as the dynamics of their transport. A reason for the lower relative mean accuracy compared to FNNs is that the movement of plumes and downwellings, while captured, can be longitudinally off. In other words, the downwelling captured in Fig. \ref{fig-lstm1} (rows 2 and 3) are slightly shifted in the angular direction. The same can be seen in all the difference plots of Fig. \ref{fig-lstm3}, at radial locations of $0.5-0.83$ for a downwelling and $0.83-1.2$ for a plume in the longitudinal center. 

One could further examine the 1D temperature profiles, obtained by horizontally averaging the 2D temperature fields (column 4 in Fig. \ref{fig-nn1}, Fig. \ref{fig-nn3}, Fig. \ref{fig-lstm1}, Fig. \ref{fig-lstm3}). LSTM temperature profiles have a mean relative absolute accuracy of $99.42\%$, while those of FNN are $99.71 \%$. While this intuitively counters the longitudinal shift argument, the mean of smudged-out/lacking plumes and downwellings predicted by the FNN can match the mean of sharp plumes and downwellings of GAIA better than the LSTM. The fact that the LSTM temperature fields capture some, but not all the downwellings in cases of vigorous convection (Fig. \ref{fig-lstm3}) can throw the horizontal mean off. In fact, finding an error metric that is invariant to, for example, longitudinal shift of a downwelling is non-trivial. \edit{\cite{Fienup97}, for example, show how modified versions of normalized root-mean-squared (NRMSE) can be computed that are invariant to certain effects such as multiplication by a constant, or phase shift for image reconstruction. Since, from a planetary evolution viewpoint, the magnitude of the temperature field matters, one could attempt to find similar shift- or rotation-invariant metrics but in terms of MSE, instead of NRMSE. Furthermore, one must consider whether to use the modified MSE expression to only evaluate the error, or also to optimize the weights of the machine learning  architectures. In the next subsection, we show that MSE seems capable of learning some non-trivial dynamics of mantle convection, as long as the underlying machine learning algorithm is suitable.} 

\begin{figure*}
\centering
\includegraphics[width=0.8\textwidth]{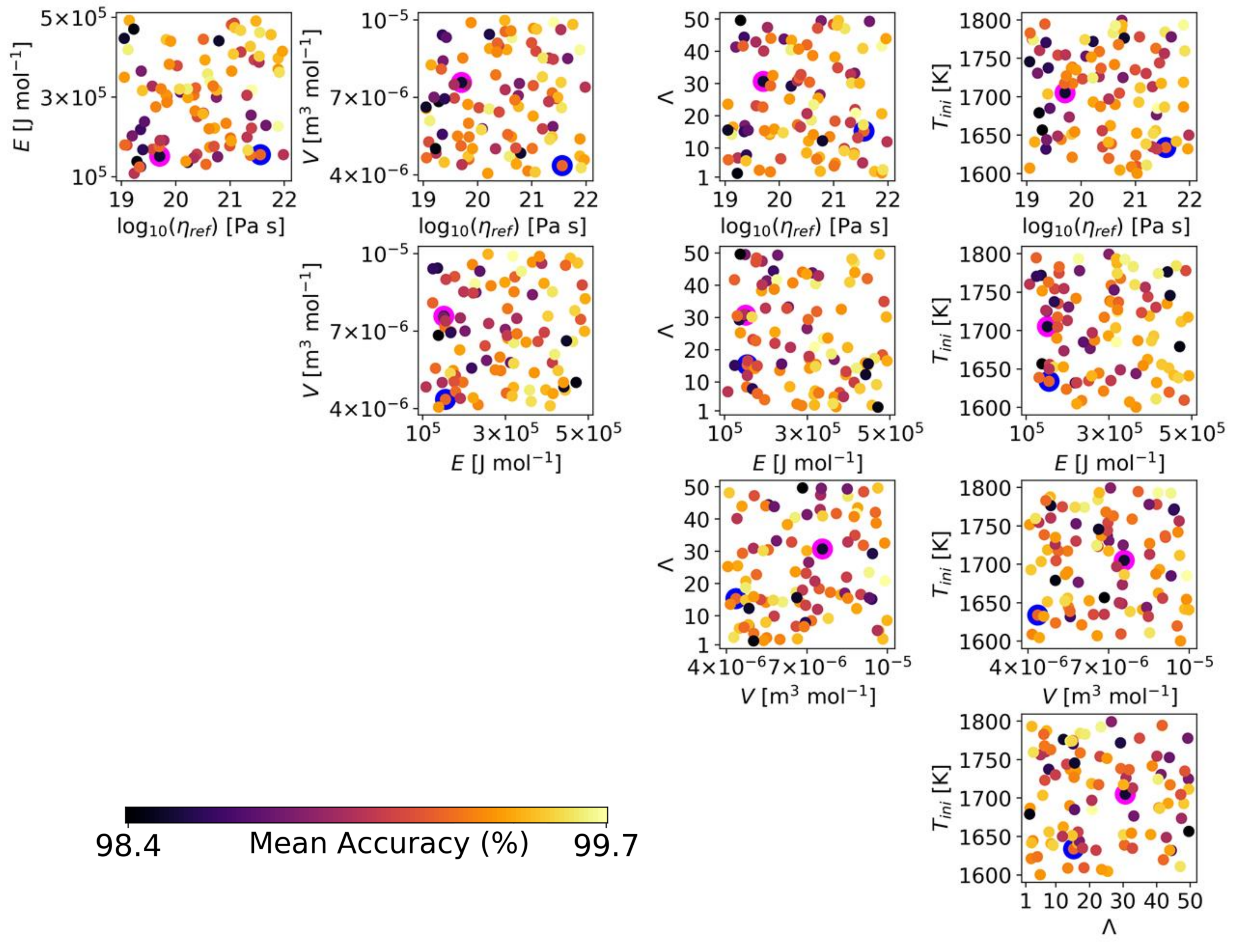}
\caption{\edit{Mean relative accuracy of LSTM predictions of the temperature fields for all the simulations in the test set with respect to the original GAIA simulations, expressed as a percentage. The mean relative accuracy is plotted with respect to two parameters at a time - indicated on the x- and y-axis with the units in which the parameters are measured.} Example in Fig. \ref{fig-lstm1} is circled in blue, while the example in Fig. \ref{fig-lstm3} is circled in magenta.}
\label{fig-lstm-all}
\end{figure*}

The \edit{mean relative accuracy} for all the simulations in the test set is provided in Fig. \ref{fig-lstm-all}. Low accuracy seems to be slightly correlated to low reference viscosity and low activation energy for the diffusion creep. A low reference viscosity generally leads to more vigorous convection, thereby inducing small-scale convection structures, which the LSTM finds difficult to predict. Similarly, a low activation energy, or equivalently, a low dependence of viscosity on temperature has the same qualitative effect of reducing viscosity. Otherwise, the method works well across the entire range of parameters.

\subsection{POD comparison of FNN and LSTM predictions}

\begin{figure}
\centering
\includegraphics[width=0.5\textwidth]{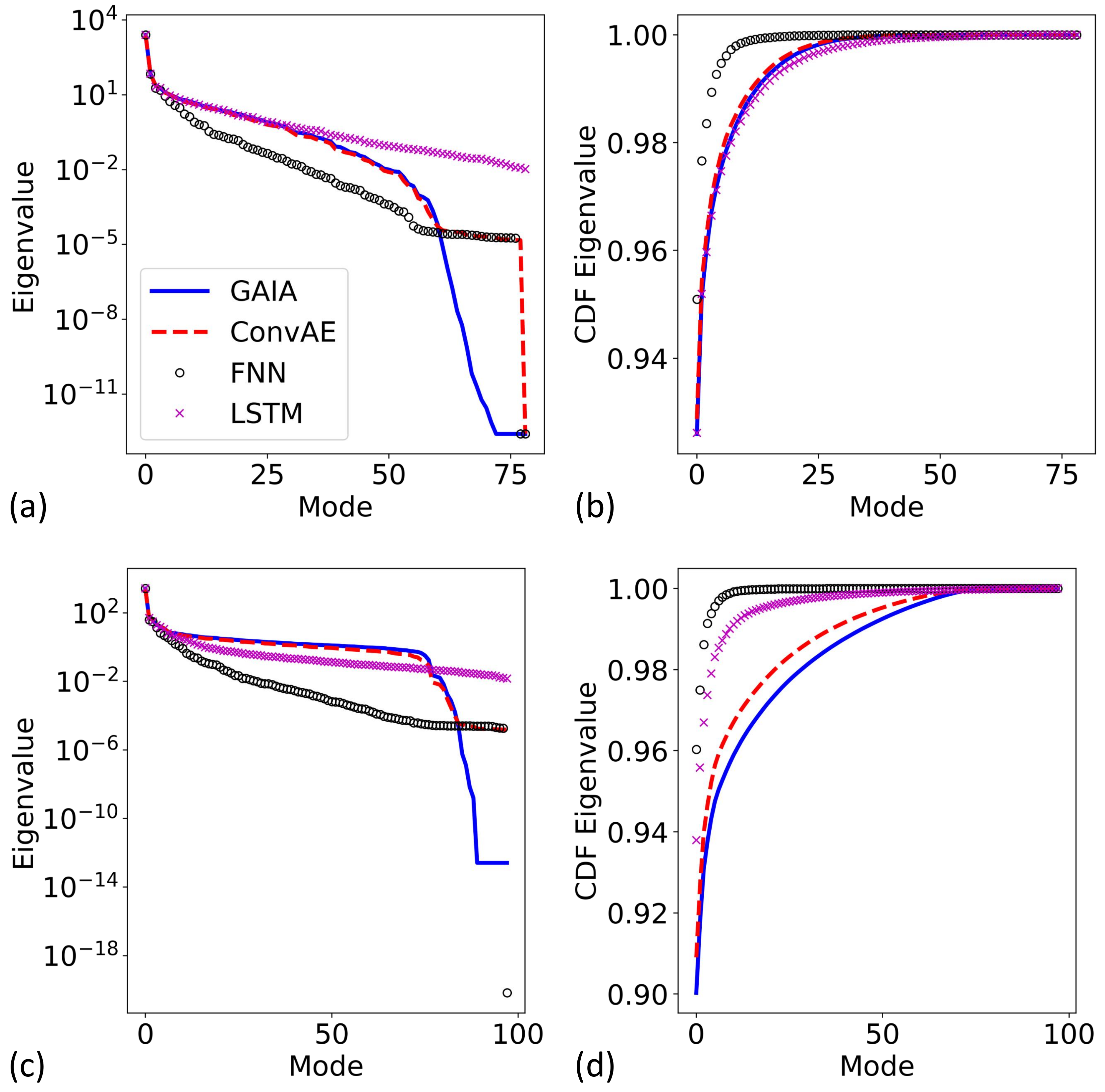}
\caption{\edit{(a) POD coefficients and (b) their cumulative distribution for example simulation 1 in the test set and (c) and (d) correspond to example simulation 2.}}
\label{fig-pod}
\end{figure}

We analyze the FNN and LSTM predictions from the lens of POD (proper orthogonal decomposition). Following e.g. \cite{BruntonKutz}, we compute the Singular Value Decomposition of a ``tall'' simulation matrix $X \in \mathbb{R}^{p \times q} $ (spatial points $\times$ time-steps):
\begin{equation}
\label{eq-pod}
X = U\Sigma V^{\ast},
\end{equation}
to obtain the spatial modes $U \in \mathbb{R}^{p \times r} $, complex conjugate $V^{\ast}$ of temporal modes $V \in \mathbb{R}^{r \times q} $ and the POD coefficients (\edit{eigenvalues}) $\Sigma \in \mathbb{R}^{r \times r} $, where $r$ is determined by the minimum of $p$ and $q$. In Fig. \ref{fig-pod}(a), we plot the \edit{eigenvalues} for the simulation in Fig. \ref{fig-lstm1} and Fig. \ref{fig-nn1} and in Fig. \ref{fig-pod}(b), we display the \edit{eigenvalues} for the simulation in Fig. \ref{fig-lstm3} and Fig. \ref{fig-nn3}.

As can be seen in both cases in Fig. \ref{fig-pod}, the \edit{eigenvalues} of an FNN-predicted temperature field evolution decay very rapidly after the first three to five modes. Hence, the cumulative distribution function (CDF) of the FNN predictions is the steepest, reaching most of its energy within the first few modes, as opposed to the other CDFs where latter modes also carry non-negligible energy. This phenomenon can be visually observed in the animations of all the five test-case simulations. See Supplemental Material at \cite{supplementary_files} for the animations. The FNN predictions are not ``energetic'' enough. The POD coefficients of the simulations predicted by LSTM decay less rapidly, even when, occasionally, the decay is desired (Fig. \ref{fig-pod}(a) modes $40$--$76$). However, in the case of vigorous convection, Fig. \ref{fig-pod}(b) shows that LSTMs, while better than FNNs, still do not fully capture the energy characteristics of the GAIA simulation. On average, the sum of \edit{eigenvalues} of the FNN predictions on the entire test set amounts to $96.51\%$ of the the sum of \edit{eigenvalues} of the GAIA predictions. For LSTM-predicted simulations, the sum of POD coefficients increases to $97.66\%$ relative to those of GAIA simulations. Thus, LSTMs capture the dynamics of the simulations better, while FNNs provide a marginally better mean accuracy for snapshots.

\section{Conclusion and future work}

We use deep learning techniques to model parameterized surrogates of mantle convection simulations. The data to train the algorithms comes from $10,525$ mantle convection simulations of a Mars-like planet, run on a 2D quarter-cylindrical grid. Focusing on only one state variable in this study - temperature - we first compressed the temperature fields using convolutional autoencoders by a factor of $142$ (from $1$ TB to $7$ GB) and then tested two regression algorithms for predicting the compressed temperature fields given five key parameters: reference viscosity (linked to the Rayleigh number), activation energy and activation volume of the diffusion creep, an enrichment factor for radiogenic elements in the crust and the initial mantle temperature (see Fig. \ref{fig-concept}(a)). 

We found that while feedforward neural networks (FNN) offer a reasonable mean accuracy, the sharper plumes and downwellings formed in the upper mantle are never transported laterally and are often lost early in the evolution. In contrast, many-to-one long-short term memory networks (LSTM) were able to capture the sharper convection structures along with their lateral transport more often, but ultimately, delivered a slightly lower mean accuracy ($99.22\%$) in comparison to FNNs ($99.30\%$). Two factors are mainly responsible for the lower mean accuracy: (1) the convection structures were longitudinally shifted and (2) the prediction error can accumulate in time. Despite this, the \edit{eigenvalues} obtained through proper orthogonal decomposition (POD) show that the FNN predictions decay too rapidly after the first three to five modes, while LSTM predictions decay less rapidly and hence capture the flow dynamics more accurately than the FNN, if not perfectly. When summed, the \edit{eigenvalues} from FNN predictions and the \edit{eigenvalues} from LSTM predictions amount to $96.51\%$ and $97.66\%$ relative to those obtained through POD of the original simulations, respectively. 

This study serves as a first-proof that deep learning can be used to model high-dimensional parameterized surrogates of mantle convection simulations. Given five parameters, the complete spatio-temporal evolution of the temperature field can be predicted up to a reasonable accuracy\edit{, i.e. the longer wavelength structures such as the 1D temperature profile and larger plumes and downwellings as well as their lateral transport can be captured, albeit not perfectly. With respect to the thermal evolution of terrestrial planets, the 2D temperature fields can be used to calculate a number of fields of interest and relate them to various quantities that can be inferred from actual observations. Lateral variations in the heat flux are important for estimating the elastic lithospheric thickness (e.g. \cite{plesa2016}). Spatio-temporal variations in the heat flux at the core-mantle boundary affect the generation and morphology of the magnetic field (e.g. \cite{amit2015}). The formation of plumes and downwellings is important for calculating the amount of melt produced during the thermal evolution and to relate this to estimates of the thickness of the crust (e.g. \cite{plesa2014b}) . The fact that plumes and downwellings are not accurately captured by the FNN can impact local melt production. Similarly, LSTM's longitudinally shifted plumes compared to the true simulations can result in slightly different looking crustal distributions. It is not straightforward to predict how consequential these errors would be to constraining the parameters. Ideally, one would conduct an inverse study to test the sensitivity of uncertainties in the observables resulting from instrumentation and/or from the surrogate model (e.g. \cite{agarwal2021}).}

Parameterized  surrogates such as the ones presented in this paper or the one proposed in \cite{duraisamy_2019} are primarily useful for performing parameter-studies - be it placing constraints on the evolution of a planet like Mars, or optimizing an airfoil to achieve the target aerodynamic performance.

This is fundamentally different from applications where the time-steps of the same simulation can be split into training and test sets (e.g., \cite{Mohan2020, pinn, DNSConvect}). On the one hand, the latter is an easier learning task because one can expect the dynamics of a single simulation to exist on a significantly smaller manifold than multiple simulations with a wide range of parameters. On the other hand, the simplicity of the flow in our simulations (e.g. no turbulence, limited compressibility and 2D flow instead of 3D) begs the question if parameterized  surrogates can be learned for more complex flows than the one used in this study.

Particularly important would be the computational cost of generating the data. For flows similar to those considered here, but in 3D, running $10,000$ simulations would be intractable. Worse yet, $10,000$ simulations could be an order of magnitude less than needed to learn spatio-temporal dynamics in 3D. \edit{This is especially important in the light of the fact that while the 2D convection models provide significantly more information than 0D (e.g., \cite{stevenson1983,gurnis1989,schubert1990,hauck2004,korenaga2011,morschhauser2011,tosi2013b,orourke2015}) or 1D evolution models \cite{agarwal2020}, they still cannot be used to constrain parameters based on localized observational constraints in 3D such as crustal thickness, elastic lithospheric thickness or surface heat flux.} Hence, \edit{this study only serves as a stepping-stone to the ultimate goal of using high-dimensional forward surrogates for probabilistic inversion of mantle convection parameters. To reach this goal,} further research is needed into data-efficient methods for parameterized surrogate modelling \edit{of 3D mantle convection}. One or both of the following approaches might hold the key. (1) One could use the partial differential equations and boundary conditions governing mantle convection to place soft and/or hard constraints on the optimization problem at hand (e.g., \cite{pinn, lu2021physicsinformed, mohan2020embedding, PhyGeoNet}). (2) Although POD bases are not appropriate for this problem because they do not generalize well among simulations with different parameters, the idea of finding a set of basis functions, that one only needs to learn the coefficients to, remains an attractive one (e.g., \cite{Kohn-Sham, GeoKernel, DNNMG}).

Despite these challenges, the combination of scientific deep learning and modern super-computing hardware presents an unprecedented opportunity for gaining robust insights into geophysical flows. 

\bigskip
A JupyterNotebook to predict the entire spatio-temporal evolution of the 2D temperature field from five parameters is available on Github: \url{https://github.com/agsiddhant/ForwardSurrogate_Mars_2D} \cite{2d_surrogate_github}.

\begin{acknowledgments}

\edit{We would like to thank the two anonymous reviewers whose comments helped improve a previous version of this paper.} We acknowledge the support of the Helmholtz Einstein International Berlin Research School in Data Science (HEIBRiDS). We also acknowledge the North-German Supercomputing Alliance (HLRN) for providing HPC resources (project id: bep00087). This work was also funded by the German Ministry for Education and Research as BIFOLD - Berlin Institute for the Foundations of Learning and Data (ref. 01IS18025A and ref 01IS18037A). We thank Klaus-Robert M{\"u}ller for the useful discussion on methods used in this paper.
\end{acknowledgments}

\bigskip

We list the author contributions following the taxonomy by \cite{acknowledgementArticle}. 
\textit{Conceptualization}: N.T., D.B., G.M.;
\textit{Methodology}: S.A., P.K., N.T.; 
\textit{Software}: S.A.; 
\textit{Validation}: S.A.; 
\textit{Investigation}: S.A.; 
\textit{Data curation}: S.A.; 
\textit{Writing–Original Draft}: S.A., N.T.; 
\textit{Writing–Review \& Editing}: S.A., N.T., P.K., D.B., G.M.;
\textit{Visualization}: S.A., N.T.;
\textit{Supervision}: N.T., D.B., P.K., G.M.; 
\textit{Funding Acquisition}: N.T., D.B., G.M.

\appendix

\section{Mathematical model of mantle  convection simulations}
\label{Section_Appendix_MC}

The mathematical model used to run the simulations is the same as the one presented in \cite{agarwal2020}. Yet, for completeness, we describe it in detail in this appendix. Numerical values of the model parameters that are shared by all simulations are listed at the end of the appendix in Table \ref{table-Mars-paras-1} and Table \ref{table-Mars-paras-2}.

\subsection{Governing equations}

The equations of conservation of mass, linear momentum and thermal energy under the extended Boussinesq approximation can be written in non-dimensional form as follows (e.g. \cite{schubert2001,king2010}):
\begin{equation}
\label{eq-EBA1}
\mathbf{\nabla}' \cdot \mathbf{u}' = 0,
\end{equation}
\begin{equation}
\label{eq-EBA2}
\begin{split}
- \mathbf{\nabla}' p' + \mathbf{\nabla}' \cdot \left [ \eta' \left( \mathbf{\nabla}' \mathbf{u}' + \left (\mathbf{\nabla}' \mathbf{u}' \right)^{\rm T} \right ) \right ]  \\ 
+ \left ( Ra \: \alpha' \: T'  - \sum_{l=1}^{3}Rb_l \: \varGamma_l \right )\mathbf{e}_r &\quad = 0,
\end{split}
\end{equation}
\begin{equation}
\label{eq-EBA3}
\begin{split}
\frac{DT'}{Dt'}- \mathbf{\nabla}' \cdot \left ( k' \mathbf{\nabla}' T' \right ) - Di \: \alpha' \left ( T' + T_0' \right )u_r' - \frac{Di}{Ra} \varPhi'  \\
- \sum_{l=1}^{3}Di\frac{Rb_l}{Ra}\frac{D\varGamma_l}{Dt}\gamma_l\left ( T' + T_0' \right )- \frac{Ra_Q}{Ra} &\quad = 0,
\end{split}
\end{equation}
Primed quantities are non-dimensional. $\mathbf{u}'$ is the velocity vector, $p'$ the dynamic pressure, $\eta'$ the viscosity, $Ra$ the thermal Rayleigh number, $\alpha'$ the thermal expansivity, $T'$ the temperature, $Rb_l$ the Rayleigh number associated with the $l$-th phase transition, $\varGamma_l$ the corresponding phase function (e.g. \cite{phase}), $\mathbf{e}_r$ the unit vector in the radial direction, $t'$ the time, $k'$ the thermal conductivity, $Di$ the dissipation number, $T_0'$ the surface temperature, $u_r'$ the radial component of the velocity, $\varPhi'$ the viscous dissipation, and $Ra_Q$ the Rayleigh number for internal heating. Viscosity, thermal expansivity and thermal conductivity are function of pressure and temperature (see eq. \eqref{eq:arrhenius}, Sec. \ref{sec_alphak}).

\subsection{Non-dimensionalization of state variables}

In eqs. \eqref{eq-EBA1}--\eqref{eq-EBA3}, the dimensional variables are scaled  as follows: 
\begin{equation}
\label{eq-EBA-ND-1}
\mathbf{u}' = \mathbf{u}\frac{\rho_{\rm m} c_{p_{\rm m}} D}{k_{\rm ref}},
\end{equation}
\begin{equation}
\label{eq-EBA-ND-2}
p' = p\frac{\rho_{\rm m} c_{p_{\rm m}} D^2}{\eta_{\rm ref} k_{\rm ref}},
\end{equation}
\begin{equation}
\label{eq-EBA-ND-3}
t' = t\frac{k_{\rm ref}}{\rho_{\rm m} c_{p_{\rm m}} D^2},
\end{equation}
\begin{equation}
\label{eq-EBA-ND-4}
T' = \frac{T-T_0}{\Delta T}.
\end{equation}
In eqs. \eqref{eq-EBA-ND-1}--\eqref{eq-EBA-ND-4}, $\rho_{\rm m}$ is the mantle density and $c_{p_{\rm m}}$ its heat capacity; $D\equiv=R_{\rm p} - R_{\rm c}$ is the mantle thickness ($R_{\rm p}$ and $R_{\rm c}$ are the planet and core radius, respectively); $k_{\rm ref}$ is the reference thermal conductivity, $\eta_{\rm ref}$ the reference viscosity (eq. \ref{eq:arrhenius}) and $\Delta T$ the initial temperature drop across the mantle. 

\subsection{Dimensionless quantities}

The following non-dimensional numbers appear in eqs. \eqref{eq-EBA1}--\eqref{eq-EBA3}:

\begin{equation}
\label{eq-Ra}
Ra = \frac{\rho_{\rm m}^2 c_{p_{\rm m}} \alpha_{\rm ref} g\Delta T D^3}{\eta_{\rm ref} k_{\rm ref}},
\end{equation}
\begin{equation}
\label{eq-para3}
Rb_l = \frac{\rho_{\rm m} c_{p_{\rm m}} \Delta \rho_l g D^3}{\eta_{\rm ref} k_{\rm ref}},
\end{equation}
\begin{equation}
\label{eq-Raq}
Ra_Q = \frac{\rho_{\rm m}^3 c_{p_{\rm m}} \alpha_{\rm ref} g H_0 D^5}{\eta_{\rm ref} k_{\rm ref}^2},
\end{equation}
and 
\begin{equation}
\label{eq-Di}
Di = \frac{\alpha_{\rm ref} g D}{c_{p_{\rm m}}},
\end{equation}
where, $\alpha_{\rm ref}$ and $k_{\rm ref}$ are the reference thermal expansivity and conductivity, $g$ is the gravitational acceleration,  $\Delta \rho_l$ is the density contrast across the $l$-th phase-transition, and $H_0$ is the initial rate of mantle heat production due to radiogenic elements.

\subsection{Thermal expansion and conductivity}\label{sec_alphak}

The temperature- and pressure-dependent thermal expansivity and conductivity are calculated using the parametrizations introduced by \cite{tosi2013a}, which in dimensional form, read:
\begin{equation}
\label{eq-EBA6}
\alpha(T,P) = \left ( a_0 + a_1 T + a_2T^{-2} \right ) \exp(-a_3 P), 
\end{equation}
\begin{equation}
\label{eq-EBA7}
k(T,P) = \left ( c_0 + c_1 P \right ) \left ( \frac{300}{T} \right )^{c_2}.
\end{equation}
Here, $a_0, \ldots, a_3$ and $c_0, \ldots, c_2$ are coefficients based on experimental  data valid for Mg-rich olivine. Numerical values of these coefficients are listed in Table \ref{table-Mars-paras-2}.

\subsection{Phase transitions}\label{sec_phasetrans}

We consider two solid-solid phase transitions in the olivine system, $\alpha$ to $\beta$-spinel and $\beta$ to $\gamma$-spinel \cite{phase}. The temperature-dependent depth of the $l$-th phase boundaries $z_l(T)$ is calculated as:
\begin{equation}
\label{eq-EBA8}
z_l(T) = z_l^0 + \gamma_l(T - T_l^0).
\end{equation}
Here, $\gamma_l$ is the Clapeyron slope (positive for both phase transitions), $z_l^0$ is the reference transition depth and $T_l^0$ the corresponding reference temperature. The phase-transition function $\varGamma_l$ is expressed in terms of $z_l$ and phase transition width $d_l$ as:
\begin{equation}
\label{eq-EBA9}
\varGamma_l = \frac{1}{2}\left ( 1 + \tanh \left( \frac{z-z_l(T)}{d_l} \right ) \right ).
\end{equation}

\subsection{Depletion of heat-producing elements and partial melting}\label{sec_melting}

We assume that a crust of thickness $d_{\rm cr}$ formed early in the planet evolution \cite{Nimmo} and that this event led to the extraction of a large amount of radiogenic elements from the mantle \cite{plesa2018}. We use an enrichment factor $\Lambda$ to modify the bulk abundance of heat-producing elements $C_0$ in the mantle (based on \cite{waenke1994}) to a new depleted composition $C_{\rm depleted}$ as follows:
\begin{equation}
\label{eq-HP1}
C_{\rm depleted} = \frac{M_m C_0}{M_{\rm cr} \left (\Lambda - 1 \right )  + M_{\rm m}},
\end{equation}
where $M_{\rm m}$ and $M_{\rm cr}$ are the mass of the mantle and crust, respectively.

We model partial melting following the approach of \cite{padovan}. Whenever the mantle temperature T locally exceeds the solidus, we calculate the melt fraction $\varphi_i$ by solving the following equation:
\begin{equation}
 \label{eq-EBA10}
 c_p \left ( T_i - T_{\rm sol} \right ) = L_{\rm m}  \varphi_i + c_p \Delta \varphi_i \Delta T_{\rm liq-sol}\left ( 1-\varphi_i \right ),
\end{equation}
where, $T_i$ is the temperature in the $i$-th cell of the computational domain, $T_{\rm sol}$ the local solidus temperature, $L_{\rm m}$ the latent heat of melting and $\Delta T_{\rm liq-sol}$ the local difference between the liquidus and solidus temperature. For the solidus and liquidus, we use the parameterization of \cite{Herzberg2000} and of \cite{Zhang1994}, respectively:
\begin{equation}
\label{eq-MeltingCurve1}
T_{\rm sol} = e_0 + e_1P + e_2 P^{2} + e_3 P^{3} + e_4 P^{4},
\end{equation}
\begin{equation}
\label{eq-MeltingCurve2}
T_{\rm liq} = f_0 + f_1 P + f_2 P^2 + f_3 P^3 + f_4 P^4,
\end{equation}
where $T_{\rm sol}$ and $T_{\rm liq}$ are the dimensional solidus and liquidus temperatures, respectively, $P$ is the dimensional hydrostatic pressure and $e_0,\ldots,e_4$ and $f_0,\ldots,f_4$ are numerical coefficients listed in Table \ref{table-Mars-paras-2}.


Using the sum of melt produced in all cells at time-step $t$, $\varphi_t$, we adjust the internal heating Rayleigh number to model the further extraction of heat-producing elements due to melting:
\begin{equation}
\label{eq-melt1}
Ra_{Q_t} = Ra_{Q_{t-1}}\left(1-\Lambda \varphi_t \right),
\end{equation}

\subsection{Evolution of the core-mantle boundary temperature}

An isothermal boundary condition is imposed at the core-mantle boundary whose temperature $T_{\rm c}$ evolves according to (e.g. \cite{stevenson1983}):
\begin{equation}
\label{eq-CoreCool}
c_{p_{\rm c}} \rho_{\rm c} V_{\rm c} \frac{\mathrm{d} T_{\rm c}}{\mathrm{d} t} = -q_{\rm c} A_{\rm c}.
\end{equation}
Here $c_{p_{\rm c}}$ is the specific heat-capacity of the core, $V_{\rm c}$ the volume of the core, $q_{\rm c}$ the average heat flux at the core-mantle boundary (CMB), and $A_{\rm c}$ the outer area of the core.


\subsection{Rescaling of the core}
\label{ref_subsection_coreRescale}
As often done in simulations of mantle convection carried out in a 2D cylindrical shell geometry, a geometric rescaling of the core radius is applied in order to better reproduce the temperature field that would be obtained in a 3D spherical shell. Following \cite{vankeken2001}, the radius of the core of the cylinder ($R_c^{\rm cyl}$) is re-scaled in such a way that the core-to-planet radius ratio is the same as the core-to-planet surface ratio of a sphere. In formulas:
\begin{equation}
\begin{split}
\label{eq-rescaleR1}
\left ( \frac{R_{\rm c}}{R_{\rm p}} \right )^2 = \frac{R_{\rm c}^{\rm cyl}}{R_{\rm p}^{\rm cyl}} \\
R_{\rm p}^{\rm cyl} + R_{\rm c}^{\rm cyl} = 1, 
\end{split}
\end{equation}
where, $R_{\rm p}$, $R_{\rm c}$ and $R_{\rm p}^{\rm cyl}$ are respectively, the radii of the spherical planet, spherical core and cylindrical planet. 

\begin{table*}
\centering
\caption{Values of fixed parameters shared by all simulations (part 1). }
\label{table-Mars-paras-1}
\begin{tabular}{llll}
\textbf{Parameter}  & \textbf{Physical meaning} & \textbf{Value}   & \textbf{Unit} \\
\hline 
\hline
$\Delta T_{t=0}$ & $^1$Initial temperature difference between core and surface & $2000$ & K  \\
$T_{\rm 0}$  & $^1$Surface temperature& $250$         & K \\
$\rho_{\rm c}$    & $^1$Core density   & $7000$   	   & kg m$^{-3}$ \\
$\rho_{\rm m}$ 	  & $^1$Mantle density & $3500$   	   & kg m$^{-3}$ \\
$c_{p_{\rm c}}$ & $^1$Core specific heat capacity & $850$  & J kg$^{-1}$ K$^{-1}$ \\
$c_{p_{\rm m}}$ & $^1$Mantle specific heat capacity & $1200$ & J kg$^{-1}$ K$^{-1}$ \\
$k_{\rm ref}$ 	& $^1$Reference thermal conductivity & $4$ & W m$^{-1}$ K$^{-1}$ \\
$\alpha_{\rm ref}$  & $^1$Reference thermal expansivity & $2.5 \times 10^{-5}$ & K$^{-1}$ \\
$R_{\rm c}$     & $^1$Outer radius of the core   & $1700$   & km \\
$R_{\rm p}$     & $^1$Planetary radius & $3400$   & km \\
$d_{\rm cr}$     & Thickness of the crust   & $64.3$   & km \\
$z_{\rm ref}$     & Reference depth for viscosity   & $232$   & km \\
$T_{\rm ref}$     & $^1$Reference temperature for viscosity   & $1600$   & K \\
\hline
$z^0_{\alpha \beta}$ &  $^1$Reference depth for $\alpha$ to $\beta$ spinel & $1020$  & km \\
$z^0_{\beta \gamma}$ &  $^1$Reference depth for $\beta$ to $\gamma$ spinel & $1360$ & km \\
$\Delta \rho^0_{\alpha \beta}$ &  $^1$Density difference for $\alpha$ to $\beta$ spinel & $250$  & kg m$^{-3}$ \\
$\Delta \rho^0_{\beta \gamma}$ & $^1$Density difference for $\beta$  to $\gamma$ spinel  & $150$  & kg m$^{-3}$ \\
$\gamma_{\alpha \beta}$ & $^1$Clapeyron slope for $\alpha$ to $\beta$ spinel  & $3 \times 10^6$  & Pa \\
$\gamma_{\beta \gamma}$ & $^1$Clapeyron slope for $\beta$ to $\gamma$ spinel  & $5.1 \times 10^6$  & Pa \\
$T_{\alpha \beta}$ &  $^1$Reference temperature for $\alpha$ to $\beta$ spinel  & $1820$  & K \\
$T_{\beta \gamma}$ & $^1$Reference temperature for $\beta$ to $\gamma$ spinel   & $1900$  & K \\
$d_l$ & $^1$Width of phase transitions  & $20$  & km \\
\hline
$^{\rm U}C_0 $  & $^2$Bulk abundance of uranium & $16\times 10^{-9}$ & kg kg$^{-1}$ \\
$^{\rm Th}C_0$  & $^2$Bulk abundance of thorium & $56\times 10^{-9}$ & kg kg$^{-1}$ \\
$^{\rm K}C_0 $ & $^2$Bulk abundance of potassium & $305\times 10^{-6}$ & kg kg$^{-1}$ \\
\hline
\multicolumn{4}{l}{$^1$\cite{plesa} $^2$\cite{Dreibus}.}
\end{tabular}
\end{table*}
\begin{table*}
\centering
\caption{Values of fixed parameters shared by all simulations (part 2).}
\label{table-Mars-paras-2}
\begin{tabular}{llll}
\textbf{Parameter}  & \textbf{Physical meaning} & \textbf{Value}   & \textbf{Unit} \\
\hline 
\hline
$a_0$ & $^{3}$Coefficient of thermal expansivity & $3.15 \times 10^{-5}$  & K$^{-1}$\\
$a_1$ & $^{3}$Coefficient of thermal expansivity & $1.02 \times 10^{-8}$  & K$^{-2}$ \\
$a_2$ & $^{3}$Coefficient of thermal expansivity & $-0.76$  & K \\
$a_3$ & $^{3}$Coefficient of thermal expansivity & $3.63 \times 10^{-2}$  & GPa$^{-1}$  \\
\hline
$c_0$ & $^{3}$Coefficient of thermal conductivity & $2.47$  & Wm$^{-1}$ K$^{-1}$  \\
$c_1$ & $^{3}$Coefficient of thermal conductivity  & $0.33$  & Wm$^{-1}$ K$^{-1}$ GPa$^{-1}$  \\
$c_2$ & $^{3}$Coefficient of thermal conductivity & $0.48$  & \\
\hline
$e_0$ & $^{4}$Coefficient for solidus parameterization & $1400$  & K \\
$e_1$ & $^{4}$Coefficient for solidus parameterization & $149.5$ & K Pa$^{-1}$ \\
$e_2$ & $^{4}$Coefficient for solidus parameterization & $-9.4$  & K Pa$^{-2}$\\
$e_3$ & $^{4}$Coefficient for solidus parameterization & $0.313$  & K Pa$^{-3}$\\
$e_4$ & $^{4}$Coefficient for solidus parameterization & $-0.0039$  & K Pa$^{-4}$\\
\hline
$f_0$ & $^{5}$Coefficient for liquidus parameterization & $1977$  & K \\
$f_1$ & $^{5}$Coefficient for liquidus parameterization & $64.1$ & K Pa$^{-1}$\\
$f_2$ & $^{5}$Coefficient for liquidus parameterization & $-3.92$  & K Pa$^{-2}$\\
$f_3$ & $^{5}$Coefficient for liquidus parameterization & $0.141$  & K Pa$^{-3}$\\
$f_4$ & $^{5}$Coefficient for liquidus parameterization & $-0.0015$  & K Pa$^{-4}$\\
\hline
\multicolumn{4}{l}{$^3$\cite{tosi2013a}  $^4$\cite{Herzberg2000} $^5$\cite{Zhang1994}.}
\end{tabular}
\end{table*}

\section{Distribution of parameters in the dataset}

\begin{figure*}
\centering
\includegraphics[width=0.55\textwidth]{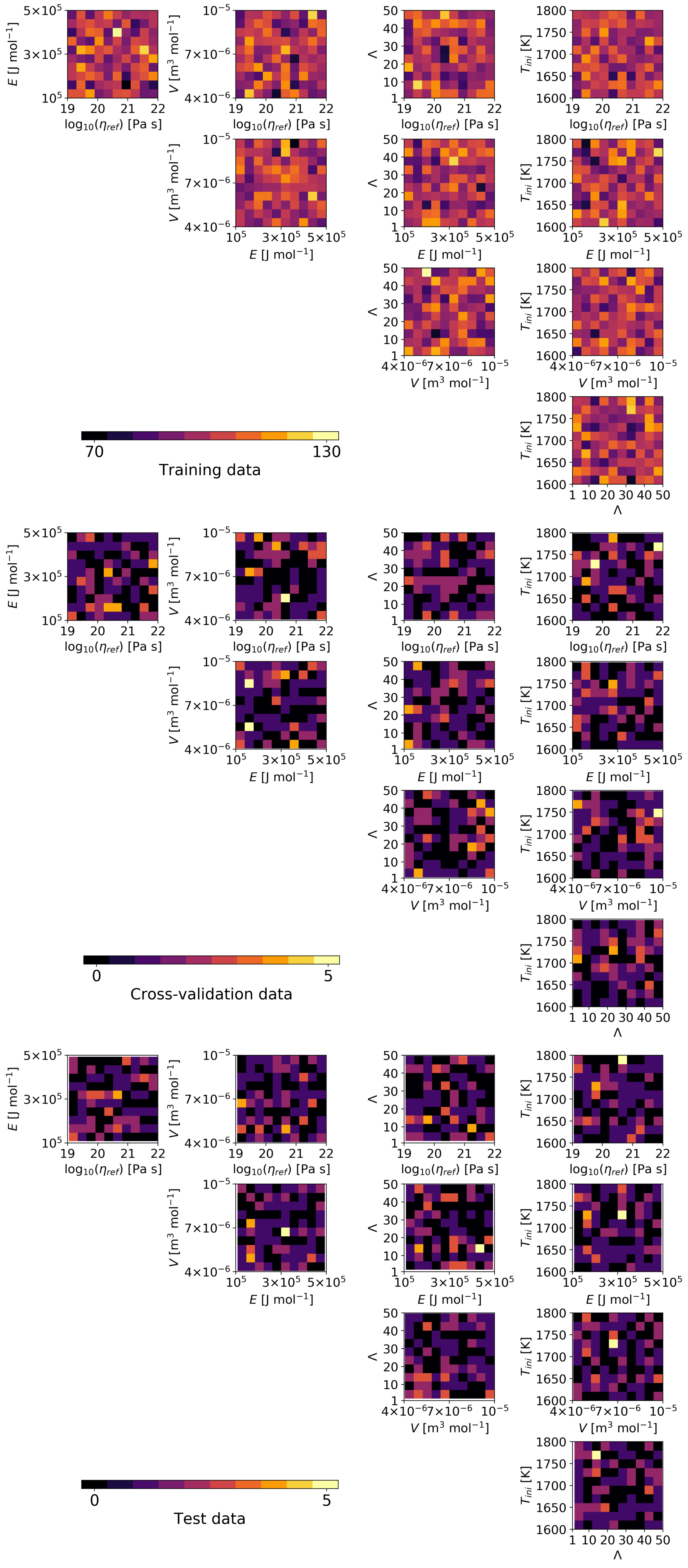}
\caption{Distribution of the simulation parameters in the training, cross-validation and test sets.}
\label{fig-paras}
\end{figure*}

\edit{Fig. \ref{fig-paras} plots the distribution of simulation parameters in the training, cross-validation and test sets. }

\section{FNN schematic}
\label{Section_Appendix_FNN}

\begin{figure*}
\centering
\includegraphics[width=0.61\textwidth]{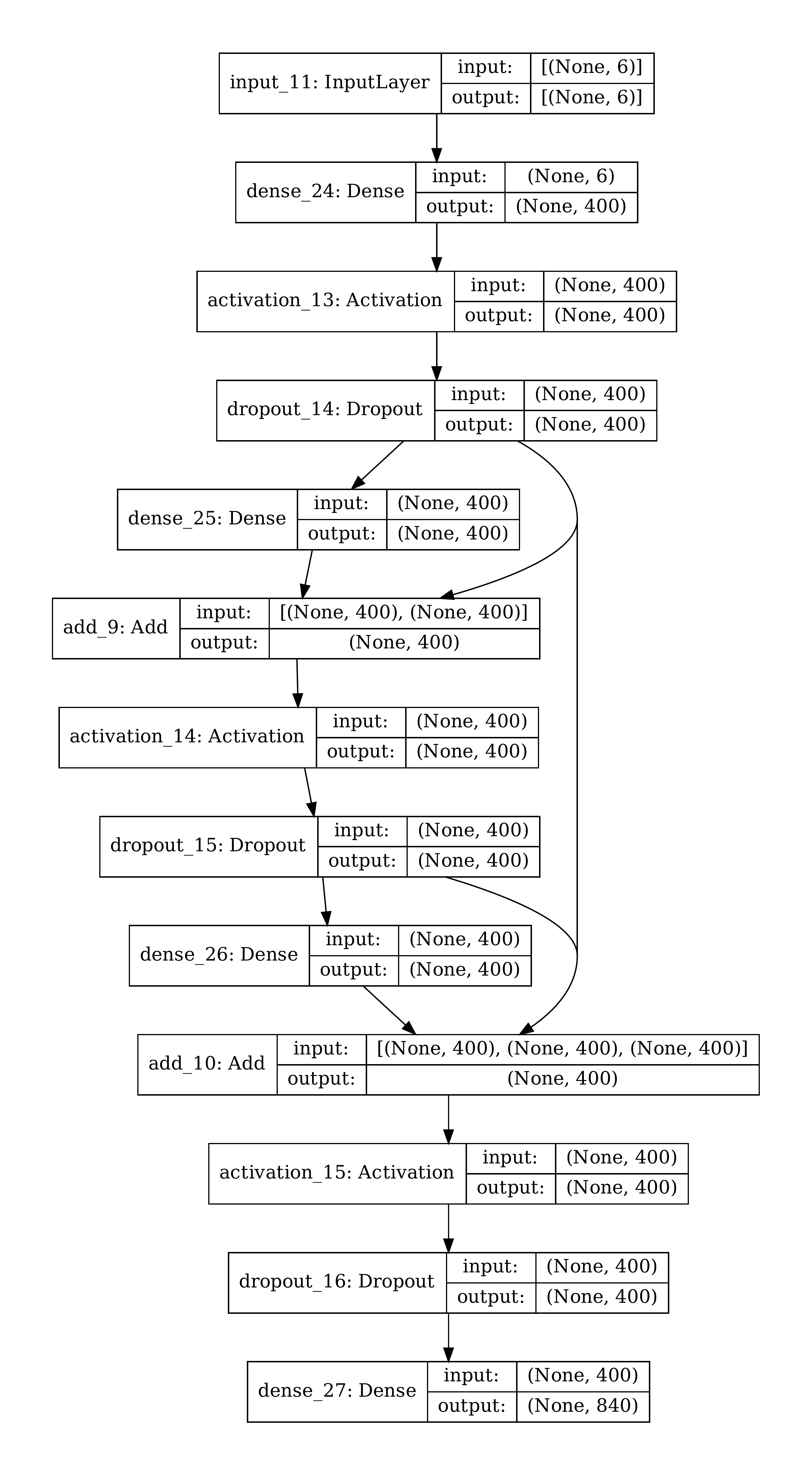}
\caption{\edit{A schematic of the FNN architecture with $3$ hidden layers with $400$ neurons each. For ease of visualization, we show only $3$ hidden layers even though the trained FNN has $8$.}}
\label{fig-fnn-schematic}
\end{figure*}

\edit{Fig. \ref{fig-fnn-schematic} is a schematic of one of the FNN architectures that were trained and the one we present the results for in Fig. \ref{fig-nn1} and Fig. \ref{fig-nn3}. The trained FNN has $8$ hidden layers with $400$ neurons each and each hidden layer is connected to all the following hidden layers via skip connections. For example, hidden layer $0$ is connected to hidden layers $1$ through $7$. For ease of visualization, we show only $3$ hidden layers in Fig. \ref{fig-fnn-schematic}.}

\section{Equations of an LSTM cell}
\label{Section_Appendix_LSTM}
Referring to Fig.  \ref{fig-lstm}(c), an LSTM cell has three main blocs. The ``forget'' gate $f_t$ determines how much information from the previous cell-state $C_{t-1}$ should be retained given the input vector $x_t$ and the previous hidden state $h_{t-1}$ (or equivalently, the previous output of an LSTM cell):
\begin{equation}
\label{eq-lstm-a}
    f_{t} = \sigma \left ( W_{f} x_{t} + U_f h_{t-1} + b_f\right ),
\end{equation}
where, $\sigma$ is the sigmoid activation ($\sigma(x) = (1/ (1+e^{-x})$), $W_f \in \mathbb{R}^{n \times m}$ is a matrix of trainable parameters, $n$ is the number of LSTM cells, $m$ is the size of the input vector $x_{t}$, $U_f \in \mathbb{R}^{n \times n}$ is another matrix of trainable parameters and $b_f \in \mathbb{R}^{n}$ is a set of biases. The subscript $f$ in $W_f$ and $b_f$ stands for the forget fate.

Then, in the ``update'' bloc, a sigmoid layer decides which values should be updated:
\begin{equation}
\label{eq-lstm-b}
    i_t = \sigma \left ( W_i x_t + U_i h_{t-1} + b_i \right ),
\end{equation}
while the $SELU$ layer creates new values $\tilde{C}_t$ to be added to the state:
\begin{equation}
\label{eq-lstm-c}
    \tilde{C}_t = SELU \left ( W_c x_t + U_c h_{t-1} + b_c \right ).
\end{equation}
$W_i$ and $W_c$ are the weights for the input connections, where subscript $i$ denotes the weights used to update values and subscript $c$ denoted the weights used to create new values. Similarly, $U_i$ and $U_c$ are weights for the recurrent connections and $b_c$ and $b_i$ are biases. 

Using Eq. \eqref{eq-lstm-a}--\eqref{eq-lstm-c}, we can now update the cell state $C_t$:
\begin{equation}
\label{eq-lstm-d}
C_t = f_t \odot C_{t-1} + i_t \odot \tilde{C}_t,
\end{equation}
where, $\odot$ is an element-wise (Hadamard) product. 

Finally, a last sigmoid layer decides the amount of cell state to be outputted via the dot product of output $o_t$ with the $SELU()$ of the cell state:

\begin{equation}
\label{eq-lstm-e}
h_t = \sigma \left ( W_o x_t + U_o h_{t-1} + b_o \right ) \odot SELU(C_t),
\end{equation}
where, $W_o$, $U_o$ and $b_o$ are the final set of trainable input weights, recurrent weights and output biases, respectively.

\bibliography{Bibliography}

\end{document}